\documentclass[10pt,journal,compsoc]{IEEEtran}

\usepackage[utf8]{inputenc} 
\usepackage[T1]{fontenc}    
\usepackage{hyperref}       
\usepackage{url}            
\usepackage{nicefrac}       
\usepackage{lipsum}

\usepackage{ifpdf}

\ifCLASSOPTIONcompsoc
  \usepackage[nocompress]{cite}
\else
  \usepackage{cite}
\fi

\ifCLASSINFOpdf
   \usepackage[pdftex]{graphicx}
\else
   \usepackage[dvips]{graphicx}
\fi

\usepackage{float}
\usepackage{epstopdf}
\usepackage{tikz-cd}
\usepackage{tabularx}

\ifCLASSOPTIONcompsoc
 \usepackage[caption=false,font=footnotesize,labelfont=sf,textfont=sf]{subfig}
\else
 \usepackage[caption=false,font=footnotesize]{subfig}
\fi

\usepackage{color}
\usepackage{xcolor}
\colorlet{edit_revision_1}{black}

\usepackage{amsthm,amsmath,amsfonts,amssymb,bm,bbm}
\newtheorem*{example*}{Example}

\newtheorem{theo}{Theorem}

\def\*#1{\bm{#1}}
\allowdisplaybreaks

\newcommand{\milp}{\texttt{milp}}
\newcommand{\bcd}{\texttt{bcd}}
\newcommand{\dprog}{\texttt{dp}}
\newcommand{\oh}{\texttt{opt-hash}}
\newcommand{\cm}{\texttt{count-min}}
\newcommand{\hh}{\texttt{heavy-hitter}}
\newcommand{\logreg}{\texttt{logreg}}
\newcommand{\cart}{\texttt{cart}}
\newcommand{\rf}{\texttt{rf}}

\makeatletter
\newcommand{\pushright}[1]{\ifmeasuring@#1\else\omit\hfill$\displaystyle#1$\fi\ignorespaces}
\newcommand{\pushleft}[1]{\ifmeasuring@#1\else\omit$\displaystyle#1$\hfill\fi\ignorespaces}
\makeatother

\usepackage[ruled,vlined,linesnumbered,resetcount]{algorithm2e}
\RestyleAlgo{boxruled}
\usepackage{algorithmic}

\begin{document}

\title{Frequency Estimation in Data Streams:\\ Learning the Optimal Hashing Scheme}

\author{Dimitris~Bertsimas 
        and~Vassilis~Digalakis~Jr.

\IEEEcompsocitemizethanks{\IEEEcompsocthanksitem D. Bertsimas is with the Sloan School of Management and the Operations Research Center, Massachusetts Institute of Technology, Cambridge,
MA, 02139.\protect\\
E-mail: dbertsim@mit.edu.
\IEEEcompsocthanksitem V. Digalakis Jr. is with the Operations Research Center, Massachusetts Institute of Technology, Cambridge,
MA, 02139.\protect\\
E-mail: vvdig@mit.edu.}
\thanks{Submitted: 07/2020. Revised: 05/2021.}}

\markboth{IEEE TRANSACTIONS ON KNOWLEDGE AND DATA ENGINEERING}%
{Bertsimas and Digalakis Jr.: Frequency Estimation in Data Streams: Learning the Optimal Hashing Scheme}
%


\IEEEtitleabstractindextext{%
\begin{abstract}
We present a novel approach for the problem of frequency estimation in data streams that is based on optimization and machine learning. 
Contrary to state-of-the-art streaming frequency estimation algorithms, which heavily rely on random hashing to maintain the frequency distribution of the data steam using limited storage, the proposed approach exploits an observed stream prefix to near-optimally hash elements and compress the target frequency distribution.
{\color{edit_revision_1}We develop an exact mixed-integer linear optimization formulation, which enables us to compute optimal or near-optimal hashing schemes for elements seen in the observed stream prefix;
then, we use machine learning to hash unseen elements.
Further, we develop an efficient block coordinate descent algorithm, which, as we empirically show, produces high quality solutions, and, in a special case, we are able to solve the proposed formulation exactly in linear time using dynamic programming.
We empirically evaluate the proposed approach both on synthetic datasets and on real-world search query data. }
We show that the proposed approach outperforms existing approaches by one to two orders of magnitude in terms of its average (per element) estimation error and by 45-90\% in terms of its expected magnitude of estimation error.
\end{abstract}

\begin{IEEEkeywords}
Data streams, streaming frequency estimation, learning to hash, optimal hashing scheme.
\end{IEEEkeywords}}

\maketitle

\IEEEdisplaynontitleabstractindextext

\IEEEpeerreviewmaketitle

\IEEEraisesectionheading{\section{Introduction}\label{sec:introduction}}
\IEEEPARstart{W}{e} consider a streaming model of computation \cite{muthukrishnan2005data,garofalakis2016data}, where the input is represented as a finite sequence of elements from some finite universe (domain) which is not available for random access, but instead arrives dynamically and one at a time in a stream. We further assume that each element is identified by a unique key and is also associated with a set of features. 
One of the most fundamental problems in the streaming model is frequency estimation, i.e., given an input stream, estimate the frequency (number of occurrences) of each element. Notice that this can trivially be computed in space equal to the minimum of the universe and the stream size, by simply maintaining a counter for each element or by storing the entire stream, respectively.
Nevertheless, data streams are typically characterized by large volume and, therefore, streaming frequency estimation algorithms should require small space, sublinear in both the universe and the stream size. Furthermore, streaming algorithms should generally be able to operate in a single pass (each element should be examined at most once in fixed arrival order) and in real-time (each element's processing time must be low). 

\begin{example*}
Consider a stream of queries arriving on a server. The universe of all elements is the set of all possible queries (of bounded length) and each element is uniquely identified by the query text. Note that any unique query may appear multiple times in the stream. The features associated with a query could include, e.g., the query length, the unigram of the query text (possibly after some pre-processing), etc. The goal is to estimate the frequency distribution of the queries, that is, the number of times each query appears in the stream, in space much smaller than the total number of unique queries.
\end{example*} 

Massive data streams appear in a variety of applications. For example, in search query monitoring, Google received more than 1.2 trillion queries in 2012 (which translates to 3.5 billion searches per day) \cite{internetlivestats}. In network traffic monitoring, AT$\&$T collects over one terabyte of NetFlow \cite{netflow} measurement data from its production network each day \cite{garofalakis2016data}. Moreover, the IPV6 protocol provides nearly $2^{128}$ addresses, making the universe of possible IP addresses gigantic, especially considering that, in many applications, we are interested in monitoring active IP network  connections between pairs (source/destination) of IP addresses. Thus, being able to process a data stream in sublinear space is essential.

Maintaining the frequency distribution of a stream of elements is useful, not only as a sufficient statistic for various empirical measures and functionals (e.g., entropy \cite{bhuvanagiri2006estimating}), but also to identify interesting patterns in the data. An example are the so-called ``heavy-hitters'' \cite{misra1982finding,cormode2009finding}, that is, the elements that appear a big number of times, which, e.g., could be indicative of denial of service attacks in network traffic monitoring (see \cite{garofalakis2016data} for a detailed discussion of applications). 

In this paper, we address the problem of frequency estimation in data streams, under the additional assumption that a prefix of the input stream has been observed. Along the lines of \cite{hsu2018learningbased}, who address the same problem and extend classical streaming frequency estimation algorithms with a machine learning component, we aim to exploit the observed prefix and the features associated with each element, and develop data-driven streaming algorithms. The proposed algorithms satisfy the small-space requirement, as they significantly compress the input frequency vector, and do operate in a single pass and in real-time, as their update and query times are constant (except for the training phase, which is more computationally demanding, since we perform optimization and machine learning).

\subsection{Streaming Frequency Estimation Algorithms}
A rich body of research has emerged in the streaming model of computation \cite{muthukrishnan2005data,garofalakis2016data}; the first streaming algorithms appeared in the early 1980s, to address, in limited space, problems such as finding the most frequently occurring elements in a stream \cite{misra1982finding}. A vast literature has since been developed, especially since the 1990s, and numerous problems, including complex machine learning tasks, such as decision tree induction \cite{domingos2000mining}, can now be solved in streaming settings.

Sketches \cite{cormode2012synopses} are among the most powerful tools to process streaming data. A sketch is a data structure which can be represented as a linear transform of the input. For example, in the context of frequency estimation, the input is the vector of frequencies (or frequency distribution) of the input elements and the sketch is computed by multiplying the frequency distribution by a fixed, ``fat'' matrix. Of course, for compactness, the matrix that performs the sketch transform is never explicitly materialized and is implicitly implemented via the use of random hash functions.

Any given sketch transform is defined for a particular task. Among the most popular sketching methods for the task of frequency estimation, are the Count-Min Sketch \cite{cormode2005improved} and the Count Sketch \cite{charikar2002finding}, which both rely on random hashing and differ in their frequency estimation procedure. Historically, the so-called AMS Sketch \cite{alon1999space}, which addresses the task of estimating the sum of the squares of the frequencies of the input stream, was among the first sketching algorithms that have been proposed. Sketching algorithms have found numerous applications, including in measuring network traffic \cite{yu2013software}, in natural language processing \cite{goyal2012sketch}, in signal processing and compressed sensing \cite{gilbert2010sparse}, and in feature selection \cite{aghazadeh2018mission}.

\subsection{Learning-Augmented Streaming Algorithms}
The abundance of data that is available today has motivated the development of the field of learning-augmented algorithms, whereby traditional algorithms are modified to leverage useful patterns in their input data. More specifically, in the context of streaming algorithms, \cite{kraska2018case} and \cite{mitzenmacher2018model} augment with a machine learning oracle the Bloom filter \cite{bloom1970space,broder2004network}, a widely used probabilistic data structure that tests set membership. \cite{hsu2018learningbased} develop learning-based versions of the Count-Min Sketch and the Count Sketch; {\color{edit_revision_1}a similar approach is taken by \cite{yang2018empowering}, who focus on network monitoring and develop a generalized framework to augment sketches with machine learning. The latter two approaches use machine learning in parallel with a standard (random) sketch, that is, they combine a machine learning oracle with standard (conventional) streaming frequency estimation algorithms, such as the Count-Min Sketch.

In this paper, we consider the same problem as in \cite{hsu2018learningbased}, namely learning-based streaming frequency estimation. However, our approach fundamentally differs in that we combine a (non-random) sketch (i.e., the optimal hashing scheme) and machine learning into a new estimator and hence our approach does not rely on random hashing at all. Instead, we use optimization to learn an optimal (or near-optimal) hashing scheme from (training) data, and machine learning to hash ``unseen elements,'' which did not appear in the training data. 
}

\subsection{Learning to Hash}

The proposed approach has connections with the field of learning to hash, a data-dependent hashing approach which aims to learn hash functions from a specific dataset (see \cite{wang2017survey} for a comprehensive survey). Learning to hash has mostly been considered in the context of nearest neighbor search, i.e., learning a hashing scheme so that the nearest neighbor search result in the hash coding space is as close as possible to the search result in the original space. Optimization-based learning to hash approaches include the works \cite{kulis2009learning,lin2013general,lin2014fast}. \cite{andoni2015optimal} develop an optimal data-dependent hashing scheme for the approximate nearest neighbor problem. {\color{edit_revision_1} To the best of our knowledge, our approach is the first that considers learning hashing schemes for the streaming frequency estimation problem, whereby the objective is different. }

\subsection{Learning-Augmented Algorithms beyond Streaming and Hashing}

Beyond streaming and hashing algorithms, \cite{purohit2018improving} use machine-learned predictions to improve the performance of online algorithms. \cite{mao2019learning} use reinforcement learning and neural networks to learn workload-specific scheduling algorithms that, e.g., aim to minimize the average job completion time. Machine learning has also been used outside the field of algorithm design, e.g., in signal processing and, specifically, in the context of ``structured'' (instead of sparse) signal recovery \cite{mousavi2015deep} and in optimization. \cite{khalil2016learning} and \cite{balcan2018learning} propose machine learning-based approaches for variable branching in mixed-integer optimization, \cite{khalil2017learning} use reinforcement learning to learn combinatorial optimization algorithms over graphs, \cite{bertsimas2018voice} use interpretable machine learning methods to learn strategies behind the optimal solutions in continuous and mixed-integer convex optimization problems as a function of their key parameters, and \cite{bertsimas2019online} focus specifically on online mixed-integer optimization problems. Machine learning has also been popularized in the context of data management and, in particular, in tasks such as learning index structures \cite{kraska2018case} and query optimization \cite{krishnan2018learning,ortiz2018learning}.

\subsection{Contributions}
Our key contributions can be summarized as follows:
\begin{itemize}
    \item[-] We develop a novel approach for the problem of frequency estimation in data streams that is based on optimization and machine learning. By exploiting an observed stream prefix, the proposed learning-based streaming frequency estimation algorithm achieves superior performance compared to conventional streaming frequency estimation algorithms.
    \item[-] We present an exact mixed-integer linear optimization formulation, as well as an efficient block coordinate descent algorithm, that enable us to compute near-optimal hashing schemes and provide a smart alternative to oblivious random hashing schemes. This part of our work could be of independent interest, beyond the problem of frequency estimation in data streams. {\color{edit_revision_1}Further, in a special case, we are able to solve the proposed formulation exactly in linear time using dynamic programming.}
    \item[-] We apply the proposed approach to the problem of search query frequency estimation and evaluate it using both synthetic and real-world data. Computational results indicate that the proposed approach notably outperforms state-of-the-art non-learning and learning-based approaches in terms of its estimation error. Moreover, the proposed approach is by construction interpretable and enables us to get additional insights into the problem of search query frequency estimation.
\end{itemize}
The rest of the paper is organized as follows.
{\color{edit_revision_1}
In Section \ref{sec:preliminaries}, we formalize the streaming frequency estimation problem and present, at a high level, the Count-Min Sketch, the most widely used random hashing-based streaming frequency estimation algorithm, and the Learned Count-Min Sketch, a learning-augmented version of the Count-Min Sketch.
Section \ref{sec:overview} gives an overview of the proposed approach.
In Section \ref{sec:learn-opt-hash}, we formulate the problem of learning the optimal hashing scheme using the observed stream prefix and develop efficient optimization algorithms.
Section \ref{sec:estimate} describes the frequency estimation procedure we apply, after the optimal hashing scheme is learned. 
In Section \ref{sec:synthetic-experiments}, we use synthetic data to explore the performance and scalability of the proposed algorithms and investigate the impact of various design choices on the proposed approach.
Section \ref{sec:experiments} empirically evaluates the proposed approach on real-world search query data. 
Section \ref{sec:conclusion} concludes the paper.
}

\section{Preliminaries}\label{sec:preliminaries}
In this section, we formally describe the problem of frequency estimation in data streams and present the state-of-the-art approaches to solving it. {\color{edit_revision_1} Although we explain the notation that we use in the main text, for convenience, we also gather the basic notations in Table \ref{tab:notations} in Appendix \ref{sec:appendix-notation}.}

Formally, we are given input data in the form of an ordered set of elements $$\mathcal{S} = (u_1 , u_2 , \dots, u_{|\mathcal{S}|}),$$ where $u_t \in \mathcal{U}, \ \forall t \in [|\mathcal{S}|]:=\{1,...,|\mathcal{S}|\}$, and $\mathcal{U}$ is the universe of input elements. Each element $u \in \mathcal{U}$ is of the form $$u = (k,\*x),$$ where (without loss of generality) $k \in [|\mathcal{U}|]$ is a unique ID and $\*x \in \mathcal{X}$ is a set of features associated with $u$. The goal is, at the end of $\mathcal{S}$, given an element $u \in \mathcal{U}$, to output an estimate $\tilde{f}_u$ of the frequency $$f_u = \sum_{t=1}^{|\mathcal{S}|} \mathbbm{1}_{(u_t=u)}$$ of that element, {\color{edit_revision_1} i.e., the number of times the element appears in $\mathcal{S}$; here, $\mathbbm{1}_{\mathcal{A}}$ denotes the indicator function of event $\mathcal{A}$}. We assume that both $\mathcal{S}$ and $\mathcal{U}$ are huge, so we wish to produce accurate estimates in space much smaller than $\min\{ |\mathcal{S}| , |\mathcal{U}| \}$. We work under the additional assumption that a prefix {\color{edit_revision_1}$\mathcal{S}_0 = (u_1 , u_2 , \dots, u_{|\mathcal{S}_0|})$ (where $|\mathcal{S}_0|\ll|\mathcal{S}|$) of the input stream has already been observed.}

\subsection{Conventional Approach: Random Sketches} The standard approach to attack this problem is the well-known Count-Min Sketch (CMS) \cite{cormode2005improved}, a probabilistic data structure based on random hashing that serves as the frequency table of $\mathcal{S}$. In short, CMS randomly hashes (via a random linear hash function $\text{hash}(\cdot)$) each element $u \in \mathcal{U}$ to a bucket in an array $\*\phi$ of size $w \ll \min\{ |\mathcal{S}| , |\mathcal{U}| \}$; whenever element $u$ occurs in $\mathcal{S}$, the corresponding counter $\*\phi_{hash(u)}$ is incremented. Since $w \ll |\mathcal{U}|$, multiple elements are mapped to the same bucket and $\*\phi_{hash(u)}$ overestimates $f_u$. {\color{edit_revision_1} In practice, multiple arrays $\*\phi^1,...,\*\phi^d$ are maintained (each array is referred to as a ``level'') and the final estimate for $f_u$ is $$\tilde{f}_u = \min_{l\in[d]} \*\phi^l_{hash^l(u)},$$ where $\text{hash}^l(\cdot)$ is the hash function that corresponds to the $l$-th level. Intuitively, by repeating the estimation procedure multiple times and taking the minimum of the estimated frequencies (all of which overestimate the actual frequency), the resulting estimator's accuracy will improve.} CMS provides probabilistic guarantees on the accuracy of its estimates, namely, for each $ u \in \mathcal{U},$ with probability $1-\delta$, $$|\tilde{f}_u - f_u| \leq \epsilon ||\mathbf{f}||_1,$$ where $\epsilon=\frac{e}{w}$ and $\delta=e^{-d}.$ In total, CMS consists of $b = w \times d$ buckets.

\subsection{Learning-Based Approach: Learned Sketches}
To leverage the observed stream prefix, \cite{hsu2018learningbased} augment the classical CMS algorithm as follows. Noticing that the elements that affect the estimation error the most are the so-called heavy-hitters (i.e., elements that appear many times), they propose to train a classifier $$h_{\text{HH}}: \mathcal{X} \rightarrow \{ \text{heavy} , \overline{\text{heavy}} \}$$ that predicts whether an element $u=(k,\*x)$ is going to be a heavy-hitter or not. 
\footnote{
{\color{edit_revision_1}
\cite{hsu2018learningbased} identify the heavy-hitters by first predicting the element frequencies (or log-frequencies) using machine learning and then selecting, using validation data, the optimal cutoff threshold for an element to be considered a heavy-hitter. In their experiments, they predict whether an item is in the top $1\%$ of the frequencies.
}}
Then, they allocate $b_{\text{heavy}}$ unique buckets to elements identified as heavy-hitters, and randomly allocate the remaining $b_{\text{random}} = b-2b_{\text{heavy}}$ buckets to the rest of the universe, using, e.g., the standard CMS. We call their algorithm the Learned Count-Min Sketch (LCMS).

An important remark is that each of the $b_{\text{heavy}}$ unique buckets allocated to heavy-hitters should maintain both the frequency and the ID of the associated element. As explained, this can be achieved by using hashing with open addressing, whereby it suffices to store IDs hashed into $\log b_{\text{heavy}}+t$ bits (instead of whole IDs which could be arbitrarily large) to ensure there is no collision with probability $1-2^{-t}$. Noticing that $\log b_{\text{heavy}}+t$ is comparable to the number of bits per counter, the space for a unique bucket is twice the space of a normal bucket. The learning augmented algorithm is shown to outperform, both theoretically and empirically, its conventional, fully-random counterpart. Additionally, they prove that under certain distributional assumptions, allocating unique buckets to heavy-hitters is asymptotically optimal \cite{hsu2018learningbased,aamand2019learned}. In general, however, their approach remains heuristic, does not guarantee optimal performance, and possibly throws away information by taking hard, binary decisions.


\section{Overview of the Proposed Approach} \label{sec:overview}
Motivated by the success of LCMS, we investigate an alternative, optimization-based approach in using the observed stream prefix to enhance the performance of the frequency estimator.

At a high level, the proposed two-phase approach works as follows. In the first phase, the elements that appeared in the stream prefix are optimally allocated to buckets based on their observed frequencies so that the frequency estimation error is minimized and, at the same time, similar elements are mapped to the same bucket. Importantly, contrary to CMS-based approaches, in the proposed approach, the estimate for an element's frequency is the average of the frequencies of all elements that are mapped to the same bucket. Therefore, we aim to assign ``similar'' elements to the same bucket. In the second phase, once we have an optimal allocation of the elements that appeared in the prefix to buckets, we train a classifier mapping elements to buckets based on their features. By doing so, we are able to provide estimates for unseen elements that did not appear in the prefix and hence their frequencies are not recorded.

The proposed hashing scheme consists of a hash table mapping IDs of elements that appeared in the prefix to buckets and the learned classifier. In addition, for each bucket, we need to maintain the sum of frequencies of all elements mapped therein. During stream processing, that is, once the estimator is ready, whenever an element that had appeared in the prefix re-appears, we increment the counter {\color{edit_revision_1}(i.e., the aggregated frequency)} of the bucket to which the element was mapped. Finally, to answer count-queries for any given element, we simply output the current average frequency of the bucket where the element is mapped (either via the hash table or via the classifier).

{\color{edit_revision_1} 
Appendix \ref{sec:appendix-flowchart} provides flowcharts for the proposed approach, which further explain the learning phase, where the stream prefix is used to learn the optimal hashing scheme and the classifier, illustrate how the proposed approach answers count queries for any input element, and show the update mechanism of the proposed approach.
}

{\color{edit_revision_1}
\section{Learning the Optimal Hashing Scheme} \label{sec:learn-opt-hash}
In this section, we develop the proposed approach in learning the optimal hashing scheme.
}

\subsection{Exact Formulation} \label{sec:mip}
Let $\mathcal{S}_0 = (u_1, ..., u_{|\mathcal{S}_0|})$ be the observed stream prefix. We denote by $f_u^0$ the empirical frequency of element $u$ in $\mathcal{S}_0$, i.e.,
$$f_u^0 = \sum_{t=1}^{|\mathcal{S}_0|} \mathbf{1}_{(u_t=u)},$$
and by $\*f^0(\mathcal{S}_0)$ the entire frequency distribution after observing $\mathcal{S}_0$.
Moreover, $\mathcal{U}_0 = \{u \in \mathcal{U}: \ f_u^0 > 0 \}$ is the set of all distinct elements that appeared in $\mathcal{S}_0$ and let $|\mathcal{U}_0| = n$.
We introduce $n\times b$ binary variables, where $b$ is the total number of available buckets, defined as
\[
    z_{ij} = \left\{
    \begin{array}{ll}
    1 , \quad \text{if } i\text{th element of } \mathcal{U}_0 \text{ is mapped to bucket } j, \\
    0 , \quad \text{otherwise.}
    \end{array}
    \right.
\]
Each row $\*z_i$ of $Z$ (where we denote $[Z]_{ij} = z_{ij})$ can be viewed as an one-hot binary hash code mapping element $i$ to one of the buckets. 
At the end of the stream and given a fixed assignment for the variables $z_{ij}$, the final estimate of the frequency of element $i \in [n]$ is 
$$
\tilde{f_i} = \sum_{j \in [b]} z_{ij}
	\left(
		\frac{\sum_{k \in [n]} z_{kj} f_k}{\sum_{k \in [n]} z_{kj}}
	\right).
$$
The resulting, e.g., absolute estimation error is $\sum_{i \in [n]} |\tilde{f_i} - f_i|$; a natural objective is to pick the variables $z_{ij}$ that minimize this absolute error in the observed stream prefix. 
An alternative objective we could pick is the expected magnitude of the absolute error $\frac{1}{\sum_{k\in [n]} f_k}\sum_{i \in [n]} f_i \cdot |\tilde{f_i} - f_i|$, whereby it is assumed that the probability $p_i$ of observing element $i$ is equal to its empirical probability in the observed stream prefix, i.e., $p_i := \frac{f_i}{\sum_{k\in [n]} f_k}$. In fact, this metric is used by \cite{hsu2018learningbased} in their theoretical analysis. However, such an approach would heavily weigh the most frequently occurring elements and would probably produce highly inaccurate estimates for less frequent elements. As we would like to achieve a uniformly small estimation error, we stick to the former objective and select the variables $z_{ij}$ that solve the optimization formulation which we will present shortly. 
{\color{edit_revision_1}
We incorporate an additional term in the objective function of the proposed formulation, to take the features associated with each element into account when computing the optimal mapping of elements to buckets. For $\lambda \in [0,1]$, we have:
\begin{equation} \label{eqn:miqp}
    \begin{split}
        \underset{Z \in \{0,1\}^{n \times b} }{\min}
        & \qquad
        \sum_{i \in [n]} 
        	\sum_{j \in [b]} z_{ij} \left[
        		\lambda \left| f_i^0 - \frac{\sum_{k \in [n]} z_{kj} f^0_k}{\sum_{k \in [n]} z_{kj}} \right| \right. \\
        & \qquad \qquad \qquad \left. +(1-\lambda) \sum_{k \in [n]} z_{kj} \| \*x_i - \*x_k \|^2
        		\right]
        	 \\
        \text{s.t.} & \qquad
        	\sum_{j \in [b]} z_{ij} = 1, \quad \forall i \in [n].
    \end{split}
\end{equation}
The parameter $\lambda \in [0,1]$ controls the trade-off between hashing schemes that map to the same bucket elements that are similar in terms of their observed frequencies in the prefix $(\lambda \rightarrow 1)$ and hashing schemes that put more weight on the elements' feature-wise similarity $(\lambda \rightarrow 0)$. Therefore, we refer to the first term in the objective as the estimation error and to the second term as the similarity error.
}

Problem \eqref{eqn:miqp} is a nonlinear binary optimization problem, so it is, in principle, hard to solve. Therefore, we next develop different approaches that can be used to solve it to optimality or near-optimality in different regimes.

\subsection{Mixed-Integer Linear Reformulation} \label{sec:milp}

As we show next, Problem \eqref{eqn:miqp} can be as reformulated as a mixed integer linear optimization problem by introducing auxiliary variables and new constraints. Formally, we have the following theorem:
\begin{theo} \label{theo:equiv}
{\color{edit_revision_1}
Problem \eqref{eqn:miqp} is equivalent with the following mixed-integer linear optimization problem:
}

\begin{equation} \label{eqn:milp}
\resizebox{\columnwidth}{!}{$ 
    \begin{split}
        \underset{\substack{Z \in \{0,1\}^{n \times b},\\ E \in \mathbb{R}_{\geq0}^{n \times b},\\ \Theta \in \mathbb{R}_{\geq0}^{n \times n \times b},\\ {\color{edit_revision_1}\Delta \in [0,1]^{n \times n \times b}}}}{\min}
        & \qquad
        {\color{edit_revision_1} \sum_{i \in [n]} \sum_{j \in [b]} \left[ \lambda \theta_{iij} + (1-\lambda) \sum_{k \in [n]} \delta_{ikj} \| \*x_i - \*x_k\|^2 \right] }\\
        \text{s.t.} 
            & \qquad \sum_{j \in [b]} z_{ij} = 1, \\
        	& \pushright{\forall i \in [n],}  \\
            & \qquad \sum_{k \in [n]} \theta_{ikj} - f_i^0 \sum_{k \in [n]} z_{kj} + \sum_{k \in [n]} f_k^0 z_{kj} \geq 0, \\
            & \pushright{\forall i \in [n], \ \forall j \in [b],} \\
        	& \qquad \sum_{k \in [n]} \theta_{ikj} + f_i^0 \sum_{k \in [n]} z_{kj} - \sum_{k \in [n]} f_k^0 z_{kj} \geq 0, \\
        	& \pushright{\forall i \in [n], \ \forall j \in [b],} \\
        	& \qquad \theta_{ikj} \geq e_{ij} - M (1-z_{kj}), \\
        	& \pushright{\forall i \in [n], \ \forall k \in [n], \ \forall j \in [b],} \\
        	& \qquad \theta_{ikj} \leq e_{ij}, \\
        	& \pushright{\forall i \in [n], \ \forall k \in [n], \ \forall j \in [b],} \\ 
        	& \qquad \theta_{ikj} \leq M z_{kj}, \\
        	& \pushright{\forall i \in [n], \ \forall k \in [n], \ \forall j \in [b],} \\
        	& \qquad {\color{edit_revision_1} \delta_{ikj} \geq z_{ij} + z_{kj} - 1,} \\
        	& \pushright{\forall i \in [n], \ \forall k \in [n], \ \forall j \in [b],} \\
        	& \qquad {\color{edit_revision_1}\delta_{ikj} \leq z_{ij},} \\
        	& \pushright{\forall i \in [n], \ \forall k \in [n], \ \forall j \in [b],} \\ 
        	& \qquad {\color{edit_revision_1}\delta_{ikj} \leq z_{kj},} \\
        	& \pushright{\forall i \in [n], \ \forall k \in [n], \ \forall j \in [b],}
        	\vspace{0.5em}
    \end{split}
    $}
\end{equation}
{\color{edit_revision_1}
where $M$ is a constant that satisfies $M \geq \max_{i \in [n]} f_i^0$.
}
\end{theo}

{\color{edit_revision_1}
\begin{proof}
The proof is presented in Appendix \ref{sec:appendix-proof}.
\end{proof}
}

Problem \eqref{eqn:milp} consists of $\mathcal{O}(n^2b)$ variables and constraints. {\color{edit_revision_1}As our computational study in Section \ref{sec:synthetic-experiments} suggests, by solving the reformulated Problem \eqref{eqn:milp}, we are able to compute optimal hashing schemes for problems with thousands of elements.} Nevertheless, solving a mixed integer linear optimization problem of that size can still be prohibitive in the applications we consider. For example, in the real-world case study in Section \ref{sec:experiments}, we map up to tens of thousands of elements to up to thousands of buckets, so Formulation \eqref{eqn:milp} would consist of variables and constraints in the order of $10^{11}$. Therefore, we next develop a tailored block coordinate descent algorithm that works well in practice. 

\subsection{Efficient Block Coordinate Descent Algorithm} \label{sec:bcd}
By exploiting the problem structure, we propose the following efficient block coordinate descent algorithm (Algorithm \ref{alg:bcd}) that can be used to either heuristically solve Problem \eqref{eqn:miqp} or compute high-quality warm starts for Problem \eqref{eqn:milp}.

\begin{algorithm*}
\caption{Block Coordinate Descent Algorithm.}
\label{alg:bcd}
\begin{algorithmic}[1]
    \REQUIRE Observed frequency vector $\*f^0 \in \mathbb{N}^{n}$, number of buckets $b \in \mathbb{N}$, hyperparameter $\lambda \in [0,1]$. 
    \ENSURE Learned one-hot hashing scheme $Z \in \{0,1\}^{n \times b}$.
    
    \vspace{0.5em}
    
    \STATE Initialize $Z$ satisfying $\sum_{j \in [b]} z_{ij} = 1, \forall i \in [n]$
    
    \STATE $\varepsilon_0 \leftarrow 0 $ \qquad \texttt{$\triangleright$ Objective function value for initial map}
    
    \FOR{$j \in [b]$}
    
        \vspace{0.5em} 
        
        \STATE \texttt{$\triangleright$ Find set of elements, cardinality, and mean for bucket $j$ in initial map:}
    
        \STATE $\mathcal{I}_j, c_j, \mu_j \leftarrow \{i \in [n]: z_{ij}=1\}, |\mathcal{I}_j|, \frac{\sum_{i\in\mathcal{I}_j}f_i^0}{c_j}$ 
        
        \vspace{0.5em}
        
        \STATE \texttt{$\triangleright$ Compute estimation error $e_j$ and similarity error $s_j$ for bucket $j$ in initial map:}
        
        \STATE $e_j, s_j \leftarrow \sum_{i \in \mathcal{I}_j} \left| f_i^0 - \mu_j \right|, \sum_{(i,k)\in\mathcal{I}_j\times\mathcal{I}_j} \| \*x_i - \*x_k \|^2$
        
        \vspace{0.5em}
        
        \STATE $\varepsilon_0 \leftarrow \varepsilon_0 + \left[ \lambda e_j + (1-\lambda)s_j \right]$ 
    
    \ENDFOR
    
    \vspace{0.5em}
    
    \STATE $t \leftarrow 0$
    
    \REPEAT
        \STATE Draw a random permutation $\*{\sigma}$ of the set $[n]$
        \FOR{$i\in[n]$}
            
            \FOR{$j\in[b]$}
            
                \vspace{0.5em}
                
                \STATE \texttt{$\triangleright$ Check if $\sigma_i$ is already in bucket $j$ and compute error with and without $\sigma_i$:}
                \IF{$\sigma_i \in \mathcal{I}_j$}
                    \STATE $\varepsilon_{\sigma_i,j} \leftarrow \lambda e_j + (1-\lambda)s_j$
                    
                    \STATE $\varepsilon_{-\sigma_i,j} \leftarrow \lambda \left( \sum_{k \in \mathcal{I}_j \setminus \{\sigma_i\}} \left| f_k^0 - \frac{c_j\mu_j - f_{\sigma_i}^0 }{c_{j}-1} \right| \right) + (1-\lambda) \left( s_j - 2 \sum_{k\in\mathcal{I}_j} \| \*x_{\sigma_i} - \*x_k \|^2 \right)$
                    
                    \STATE \texttt{$\triangleright$ Update bucket $j$ stats and errors after removing $\sigma_i$:}
                    
                    \STATE $\mathcal{I}_j, c_j, \mu_j \leftarrow \mathcal{I}_j \setminus \{\sigma_i\}, c_j-1, \frac{c_j\mu_j - f_{\sigma_i}^0 }{c_{j}-1}$
                    
                    \STATE $e_j, s_j \leftarrow \sum_{k \in \mathcal{I}_j} \left| f_k^0 - \mu_j \right|, s_j - 2 \sum_{k\in\mathcal{I}_j} \| \*x_{\sigma_i} - \*x_k \|^2$
                    
                \ELSE 
                    \STATE $\varepsilon_{\sigma_i,j} \leftarrow \lambda \left( \sum_{k \in \mathcal{I}_j \cup \{\sigma_i\}} \left| f_k^0 - \frac{c_j\mu_j + f_{\sigma_i}^0 }{c_{j}+1} \right| \right) + (1-\lambda) \left( s_j + 2 \sum_{k\in\mathcal{I}_j} \| \*x_{\sigma_i} - \*x_k \|^2 \right)$
                    
                    \STATE $\varepsilon_{-\sigma_i,j} \leftarrow \lambda e_j + (1-\lambda)s_j$
                    
                \ENDIF
                
                \vspace{0.5em}
                
            \ENDFOR
            
            \vspace{0.5em}
                
            \STATE \texttt{$\triangleright$ Find best bucket $j^{\star}$ and update stats and errors after mapping $\sigma_i$ to it:}
            
            \STATE $j^{\star} \leftarrow \text{argmin}_{j \in [b]} \varepsilon_{\sigma_{i},j} + \sum_{\ell \in [b] \setminus \{j\}} \varepsilon_{-\sigma_{i},\ell} $
            
            \STATE $\*z_i \leftarrow \*e_{j^{\star}}$ \qquad \texttt{$\triangleright$ $\*e_{j^{\star}}$ denotes the $j^{\star}$-th standard unit vector}
            
            \STATE $\mathcal{I}_{j^{\star}}, c_{j^{\star}}, \mu_{j^{\star}} \leftarrow \mathcal{I}_{j^{\star}} \cup \{\sigma_i\}, c_{j^{\star}}+1, \frac{c_{j^{\star}}\mu_{j^{\star}} + f_{\sigma_i}^0 }{c_{j^{\star}}+1}$
                
            \STATE $e_{j^{\star}}, s_{j^{\star}} \leftarrow \sum_{k \in \mathcal{I}_{j^{\star}}} \left| f_k^0 - \mu_{j^{\star}} \right|, s_{j^{\star}} + \sum_{k\in\mathcal{I}_{j^{\star}}} \| \*x_{\sigma_i} - \*x_k \|^2$
            
        \ENDFOR
        
        \STATE $t \leftarrow t + 1$
        \STATE $\varepsilon_t \leftarrow \sum_{j \in [b]} \left[ \lambda e_j + (1-\lambda)s_j \right]$ 
        
    \UNTIL{$\varepsilon_{t-1}-\varepsilon_t < \epsilon$}
    \RETURN $Z$ 
\end{algorithmic}
\end{algorithm*}

Concerning the algorithm's initialization, we start from a random allocation of elements to buckets. Alternatively, we could sort elements in $\mathcal{U}_0$ in terms of their observed frequencies and allocate the first $\left[ \frac{\mathcal{U}_0}{b} \right]$ elements to the first bucket, the next $\left[ \frac{\mathcal{U}_0}{b} \right]$ to the second bucket, and so forth, or we could even use the heavy-hitter heuristic (that is, assign heavy-hitters to their own bucket and the remaining elements at random). 

{\color{edit_revision_1}In our implementation, we maintain, for each bucket, the set of elements $\mathcal{I}_j$ mapped therein, its cardinality $c_j$ and mean frequency $\mu_j$, as well as the associated estimation error $e_j = \sum_{i \in \mathcal{I}_j} \left| f_i^0 - \mu_j \right|$ and similarity error $s_j = \sum_{(i,k)\in\mathcal{I}_j\times\mathcal{I}_j} \| \*x_i - \*x_k \|^2$. After any update performed by Algorithm \ref{alg:bcd} we only need to update the above quantities, instead of having to recompute them from scratch and, therefore, we can directly evaluate the objective function value $\varepsilon$ associated with any particular mapping of elements to buckets.}

In each iteration, Algorithm \ref{alg:bcd} examines sequentially and in random order all $n$ blocks of $b$ variables $\*z_i, \ i\in[n]$. Notice that each block contains all possible mappings of a particular element to any bucket. For each element $i$, we greedily select the mapping that minimizes the overall estimation error. To do so, we remove element $i$ from its current bucket and compute the estimation error associated with each bucket $j$, first with element $i$ allocated to bucket $j$ and then without element $i$. We allocate element $i$ to the bucket $j^{\star}$ that minimizes the sum of all error terms.

The algorithm terminates when the improvement in estimation error is negligible; in case we are willing to obtain an intermediate solution faster, the termination criterion can be set to a user-specified maximum number of iterations. As we empirically show, Algorithm \ref{alg:bcd} converges to a local optimum after a few tens of iterations and produces high-quality solutions. Given that algorithm is not guaranteed to converge to a globally optimum solution, the process can be repeated multiple times from different starting points. 

{\color{edit_revision_1}
Algorithm \ref{alg:bcd} can be efficiently implemented so that the complexity of each iteration is $\mathcal{O}(n^2b)$. This is to be expected since, for each bucket, we need to compute the similarity error between all pairs of elements mapped therein, which requires $\mathcal{O}(n^2b)$ operations. 
}

{\color{edit_revision_1}

\subsection{The $\lambda=1$ Case: Efficient Dynamic Programming Algorithm} \label{sec:dp}
In the special case where we set $\lambda=1$, that is, we do not take the features into account when computing the optimal hashing scheme, we obtain the following formulation:

\begin{equation} \label{eqn:miqp-simple}
    \begin{split}
        \underset{Z \in \{0,1\}^{n \times b} }{\min}
        & \qquad
        \sum_{i \in [n]} 
        	\sum_{j \in [b]} z_{ij}
        		\left| f_i^0 - \frac{\sum_{k \in [n]} z_{kj} f^0_k}{\sum_{k \in [n]} z_{kj}} \right|
        	 \\
        \text{s.t.} & \qquad
        	\sum_{j \in [b]} z_{ij} = 1, \quad \forall i \in [n].
    \end{split}
\end{equation}

Problem \eqref{eqn:miqp-simple} is an one-dimensional k-median clustering problem and has been thoroughly studied in the literature. It is fairly straightforward to develop an $O(n^2b)$ dynamic programming algorithm to solve Problem \eqref{eqn:miqp-simple} to provable optimality as per \cite{wang2011ckmeans}. An even more efficient solution method for Problem \eqref{eqn:miqp-simple} has been developed in the context of optimal quantization; using dynamic programming in combination with a matrix searching technique, \cite{wu1991optimal} solves Problem \eqref{eqn:miqp-simple} to optimality in $O(nb)$ time. We refer the interested reader to \cite{gronlund2017fast} for a detailed and unified presentation of the above methods.

Given that we can obtain an optimal solution to Problem \eqref{eqn:miqp-simple} very fast, in $O(nb)$ time, we propose to use it as a warm start for the general $\lambda \in [0,1)$ case. Therefore, we provide another alternative for the initialization step of Algorithm \ref{alg:bcd}, in addition to the ones discussed in Section \ref{sec:bcd}.
}

\section{Frequency Estimation} \label{sec:estimate}
In this section, we describe the frequency estimation component of the proposed estimator, which, in its simplest form, consists of a multi-class classifier.

\subsection{Frequency Estimation for Elements Seen in the Prefix}
Once the optimal assignment $Z$ is computed, we essentially have a hash code $h_i = \sum_{j \in [b]} j \cdot \mathbbm{1}_{(z_{ij}=1)}, \ i\in[n],$ for each element $u \in \mathcal{U}_0$. Therefore, for element $u \in \mathcal{U}_0$, indexed by $i\in[n]$, we simply estimate its frequency as $$ \tilde{f_i} = \frac{\sum_{k \in [n]: h_k=h_i} f_k}{\sum_{k \in [n]: h_k=h_i} 1} = \mu_j.$$
We denote by $h_{\text{S}}: \mathcal{U}_0 \rightarrow [b]$ the function that maps elements seen in the prefix to buckets according to the learned hash code.

\subsection{Similarity-Based Frequency Estimation for Unseen Elements} \label{sec:similarity-based-estimation}
To be able to produce frequency estimates for elements that did not appear in the prefix, i.e., $u \in \mathcal{U} \setminus \mathcal{U}_0$, we formulate a multi-class classification problem, mapping elements to buckets based on their features. Formally, we search for a function $$h_{\text{U}}: \mathcal{X} \rightarrow [b].$$ The training set consists of all data points in $$\{(\*x_i,h_i): u_i=(k_i,\*x_i) \in \mathcal{U}_0\},$$ that is, all feature-hash code tuples for elements that appeared in the prefix. Such a classifier will allow us to estimate the frequencies of unseen elements based on the average of the frequencies of elements that ``look'' similar. The estimate for element $u = (k,\*x) \in \mathcal{U} \setminus \mathcal{U}_0$ is then
$$ \tilde{f_u} = \frac{\underset{k \in [n]: \\ h_k=h_{\text{U}}(x)}{\sum}f_k}{\underset{k \in [n]: \\ h_k=h_{\text{U}}(x)}{\sum}1}.$$

\subsection{Adaptive Counting Extension: Keeping Track of the Frequencies of Unseen Elements} \label{sec:adaptive-counting-estimation}
So far, we have described a static approach; we learn the optimal hashing scheme for the elements that appear in the stream prefix and then keep track only of their frequencies. The estimated frequencies for all elements are based only on the frequencies of elements in $\mathcal{U}_0$ (which appeared in $S_0$). We next describe a dynamic approach, that keeps track of the frequencies of elements beyond the ones in $\mathcal{U}_0$. At a high level, the adaptive approach is based on approximately counting the distinct elements in each bucket. We work as follows.
\begin{enumerate}
    \item We learn the optimal hashing scheme based on the observed stream prefix and train a classifier mapping elements to buckets, as outlined above. For each bucket, we only record the number of elements that are mapped therein (instead of storing the IDs of the elements that are mapped to this bucket). We use the classifier to determine which bucket any element is mapped to.
    \item We maintain a Bloom filter \cite{bloom1970space} $\text{BF}$, i.e., a probabilistic data structure that, given a universe of elements $\mathcal{U}$ and a set $\mathcal{U}'\subseteq\mathcal{U}$, probabilistically tests, for any element $u \in \mathcal{U}$, whether $u \in \mathcal{U}'$ (here, $\mathcal{U}'$ corresponds to the elements that have appeared in the stream). If $u \in \mathcal{U}'$, then we deterministically have that $\text{BF}(u)=1$. However, if $u \not\in \mathcal{U}'$, then it need not be the case that $\text{BF}(u)=0$ (therefore a Bloom filter is prone to false positives - we will explain the impact of those in the sequel).
    \item We initialize the Bloom filter based on the elements $u\in \mathcal{U}_0$. Therefore, all elements $u\in\mathcal{U}_0$ will initially have $\text{BF}(u)=1$. On the other hand, elements $u\not\in\mathcal{U}_0$ may initially have either $\text{BF}(u)=0$ or $\text{BF}(u)=1$.
    \item For every subsequent element $u$ that appears in the stream after the stream prefix $S_0$ has been processed, we map it to a bucket $j \in [b]$ using the trained classifier. Then, we test whether we have already seen $u$, using the Bloom filter. If $\text{BF}(u)=0$, we increase both the frequency $\phi_j$ and the number of elements $c_j$ in the bucket $j$, and we set $\text{BF}(u)=1$. If $\text{BF}(u)=1$, we only increase the frequency $\phi_j$.
    \item When queried for the frequency of any element $u \in \mathcal{U}$, regardless of whether it appeared in $\mathcal{U}_0$ or not, we estimate $$ \tilde{f_u} = \frac{\phi_j}{c_j}BF(u),$$ where $j$ is the bucket in which $u$ is mapped using the classifier.
\end{enumerate}

{\color{edit_revision_1}
The impact of Bloom filters' false positives is that the proposed approach will mark as seen elements that have not appeared in the stream. When one such element actually appears in the stream, we will not increase the counter $c_j$ that tracks the number of elements in the bucket $j$ where this element is mapped. Therefore, the estimated number of elements $c_j$ in bucket $j$ will be less than the actual number. As a result, the adaptive counting extension will generally overestimate elements' frequencies.

The flowchart for the adaptive counting extension of the proposed approach is given in Figure \ref{fig:update-bloom} in Appendix \ref{sec:appendix-flowchart}
}

{\color{edit_revision_1}

\section{Experiments on Synthetic Data}\label{sec:synthetic-experiments}
In this section, we empirically evaluate the proposed approach on synthetic data. We investigate the performance and scalability of the optimization approaches discussed in Section \ref{sec:learn-opt-hash}, and explore the possibility of using different classifiers for unseen elements (as per Section \ref{sec:estimate}).

\subsection{Data Generation Methodology}
The data that we use in our synthetic experiments are generated according to the following methodology:
\begin{itemize}
    \item[-] Elements: We parameterize the universe of elements $\mathcal{U}$ by a positive integer $G \in \mathbb{Z}_{>0}$ that controls the problem size in the way that we explain next. We generate $G$ groups of elements $\mathcal{G}_1,\dots,\mathcal{G}_G$ of exponentially increasing sizes $2^{G_0+1},\dots,2^{G_0+G}$ (where $G_0 \in \mathbb{Z}_{\geq0}$ is an additional parameter that determines the size of the smallest group; we use $G_0 = 2$ in our experiments). We associate each group $\mathcal{G}_g, g\in[G],$ with a $p$-dimensional normal distribution (we use $p=2$ in our experiments to enable visualization) with mean $\*\mu_g$ selected uniformly at random from $[-10,10]^p$ and covariance matrix equal to the identity. We draw the features associated with each element $u \in \mathcal{G}_g$ as a realization of the $p$-dimensional normal distribution $\mathcal{N}(\*\mu_g,I)$ that corresponds to the element's group.
    
    \item[-] Stream: We generate the data stream $\mathcal{S}$ according to the following process. We associate each group $\mathcal{G}_g, g\in[G]$ with an arrival probability that is proportional to $\frac{1}{g}$. Within group $\mathcal{G}_g$, we assign to each element $u\in\mathcal{G}_g$ a uniform probability of arrival $\frac{1}{|\mathcal{G}_g|}$. Thus, smaller groups are more likely to appear and elements therein have a larger probability of selection so that they represent the heavy hitters. We construct the stream by first selecting the group that each new arrival belongs to and then selecting the actual element from within that group. As far as the stream prefix $\mathcal{S}_0$ is concerned, we would want to mimic a real-world scenario where not all elements from within each group start appearing since the beginning of the stream. Therefore, when we generate the prefix, we only allow for a fraction $g_0\in [0,1]$ of elements to be selected from within each group $\mathcal{G}_g, g\in[G],$ each with probability $\frac{1}{g_0|\mathcal{G}_g|}$. Finally, we remark that, in our experiment, we generate a stream prefix of size $|\mathcal{S}_0| = 10 \cdot 2^{G}$.
    
\end{itemize}
For example, by setting $G=10$ and $g_0=0.5$, we obtain a problem with $8,192$ elements, out of which we only allow for $4,096$ to appear in the prefix, which in turn has size $10,240$. Therefore, we aim to learn a hashing scheme that maps at most $4,096$ elements to $10$ buckets; the memory requirements of such a hashing scheme would be $\approx 20$ KB.

\subsection{Algorithms and Software}
We next summarize the algorithms and software that we use in our experiments. We note that all algorithms were implemented in Python 3 and all experiments were performed on a standard Intel(R) Xeon(R) CPU E5-2690 @ 2.90GHz running CentOS release 7. We independently repeat each experiment $10$ times and report the averaged error, as well as its standard deviation.

We implement and refer to the optimization algorithms presented in Section \ref{sec:learn-opt-hash} as follows:
\begin{itemize}
    \item[-] \milp: Solves the mixed-integer linear optimization problem (Problem \eqref{eqn:milp}) from Section \ref{sec:milp} using the commercial MIO solver Gurobi \cite{gurobi2016gurobi}.
    \item[-] \bcd: Implements the block coordinate descent algorithm (Algorithm \ref{alg:bcd}) from Section \ref{sec:bcd}.
    \item[-] \dprog: Solves Problem \eqref{eqn:miqp-simple} in linear time via dynamic programming (Section \ref{sec:dp}) using a Python wrapper for the R package Ckmeans.1d.dp \cite{wang2011ckmeans}.
\end{itemize}

The machine learning algorithms that we examine include a linear classifier, namely, multinomial logistic regression (\logreg), a tree-based classifier, namely, CART (\cart) \cite{breiman1984classification}, and an ensemble classifier, namely, random forest (\rf) \cite{breiman2001random}. All methods are tuned using $10$-fold cross validation; the hyperparameters that we tune are the weight of a ridge regularization term for \logreg, the minimum impurity decrease and the maximum depth for \cart, the maximum number of features in each split and the maximum depth for \rf. Unless stated otherwise, we use \cart\ as the underlying classifier in our experiments. We use the Scikit-learn machine learning package's implementation of all the above algorithms \cite{scikit-learn}. 

Finally, we use the following notation for the frequency estimation algorithms presented in this paper. We refer to the proposed estimator as \oh. We refer to CMS (the standard Count-Min Sketch) as \cm\ and to LCMS (the learned Count-Min Sketch with the heavy-hitter heuristic) as \hh. We implement all the above estimators in Python.

\subsection{Visualization: Learned Hash Code for Seen and Unseen Elements}

In Figure \ref{fig:vis}, we show an instance of a synthetically generated problem with $G=10$ groups (Figure \ref{fig:vis_true} colors elements depending on their actual group). Figure \ref{fig:vis_freqs} shows the logarithm of the frequency of each element that appeared in a prefix of length $ |\mathcal{S}_0| = 1,000$; we assume that a fraction of $g_0=0.33$ elements from each group can appear in the prefix. In Figure \ref{fig:vis_prefix}, we present the learned hash code for elements that actually appeared in the prefix (using the \bcd\ algorithm), whereas Figure \ref{fig:vis_unseen} illustrates the hash code predicted for unseen elements (using \cart).

\begin{figure*}[!ht]
\centering
\subfloat[Element groups. ]{\includegraphics[width=0.48\columnwidth]{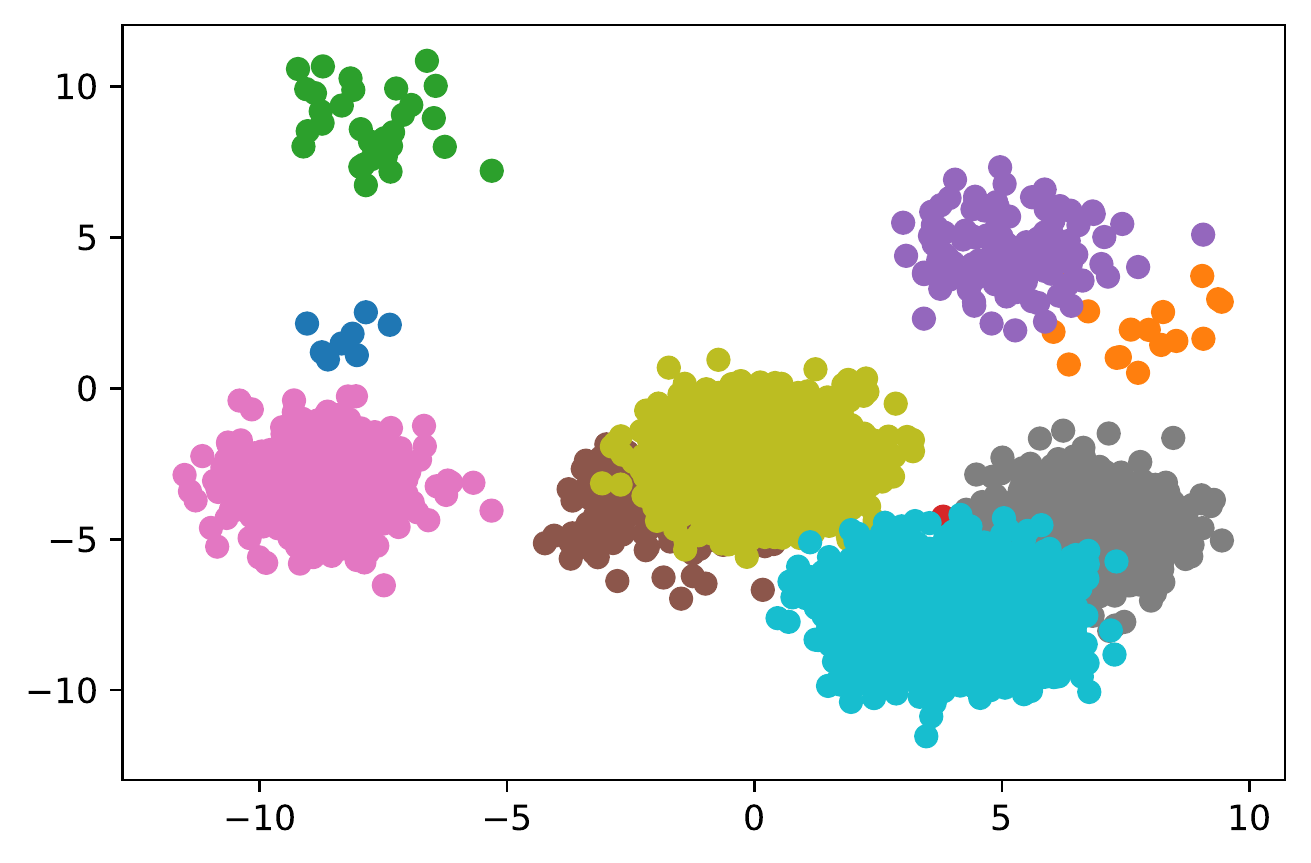}%
\label{fig:vis_true}}
\hfil
\subfloat[Prefix element frequencies. ]{\includegraphics[width=0.48\columnwidth]{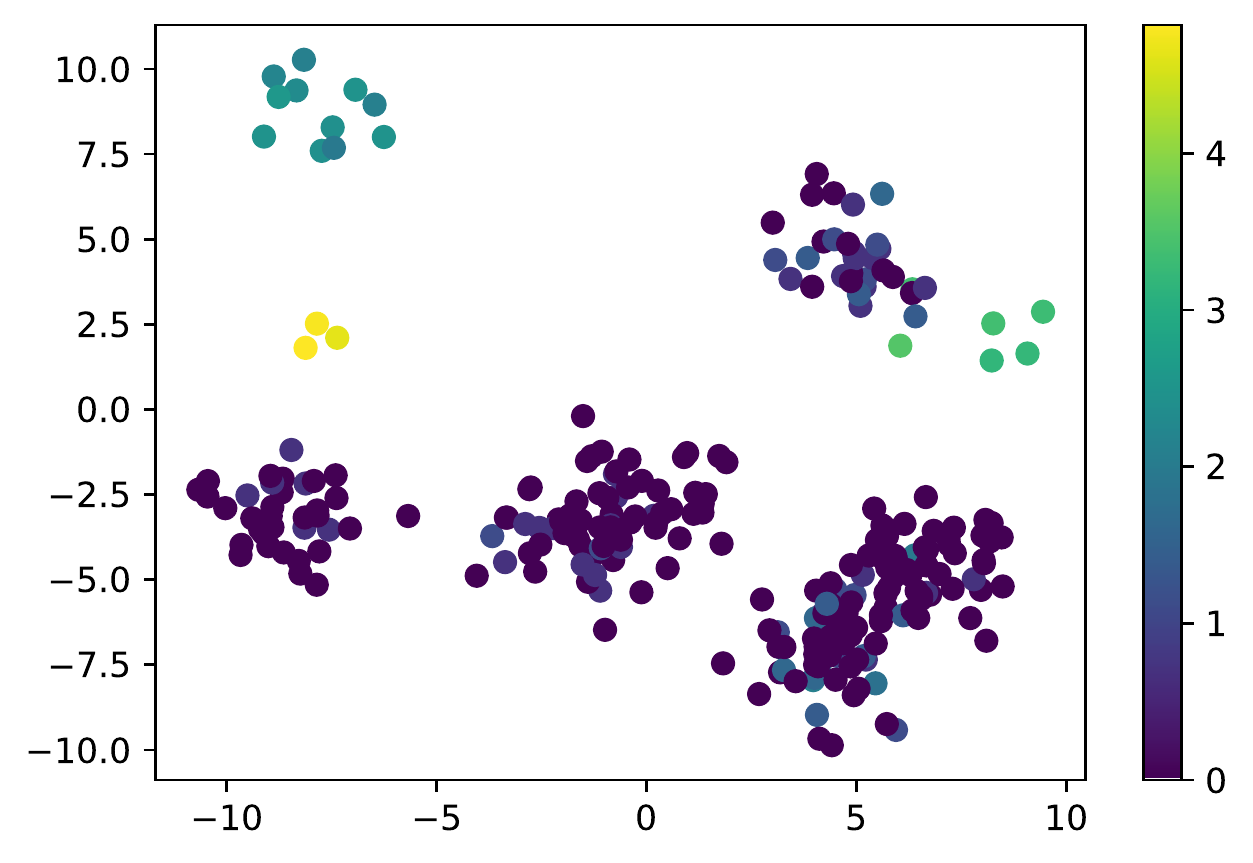}%
\label{fig:vis_freqs}}
\hfil
\subfloat[Hash code for elements that appeared in the prefix. ]{\includegraphics[width=0.48\columnwidth]{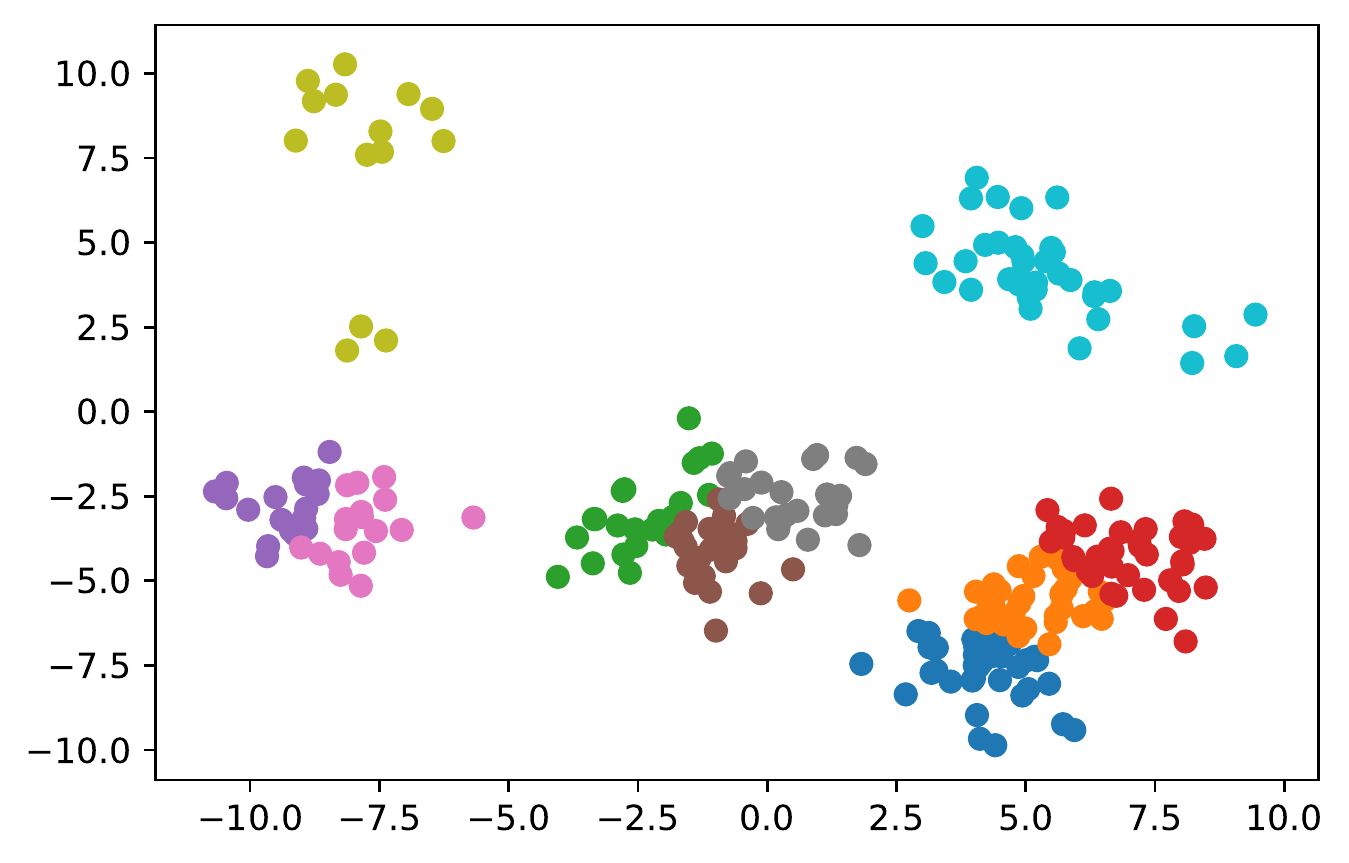}%
\label{fig:vis_prefix}}
\hfil
\subfloat[Hash code for unseen elements. ]{\includegraphics[width=0.48\columnwidth]{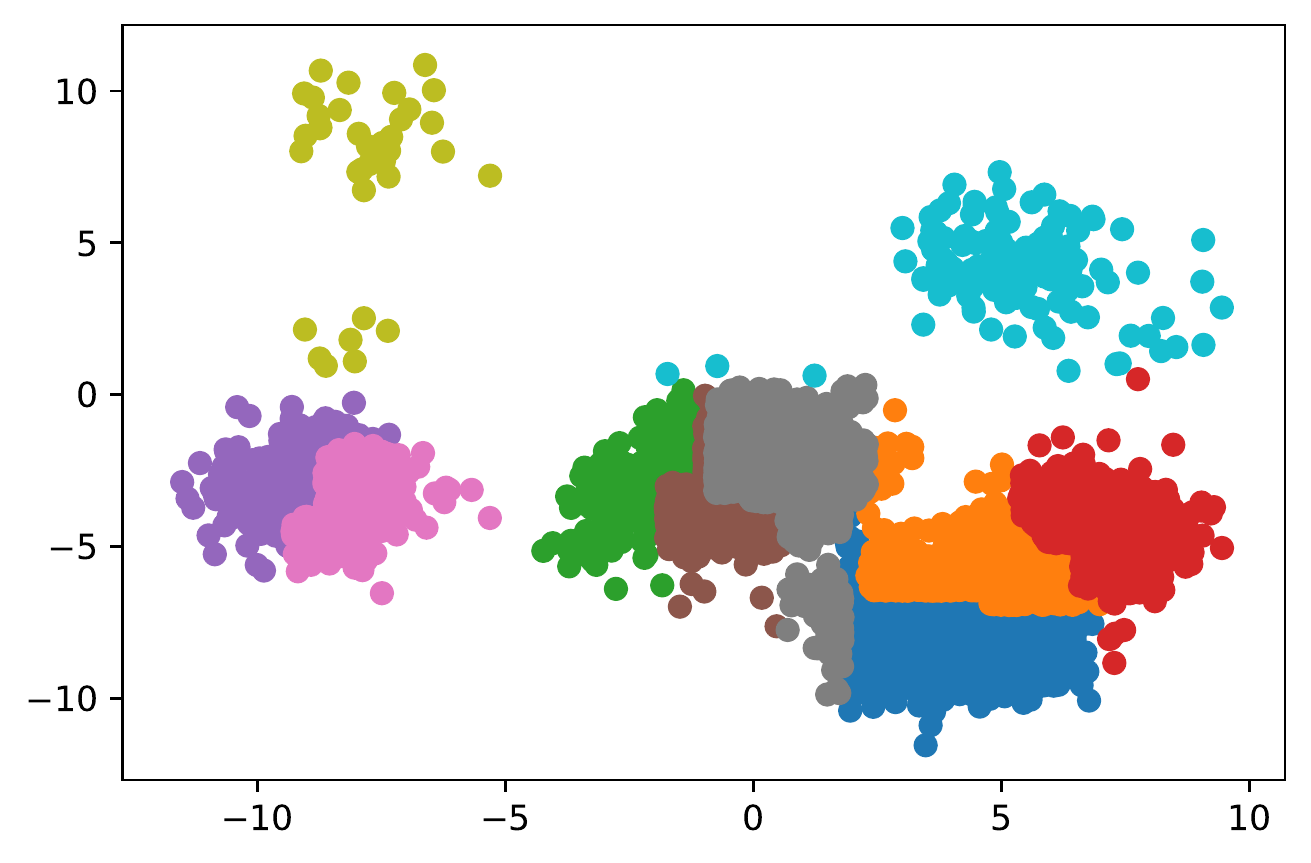}%
\label{fig:vis_unseen}}
\hfil
\caption{Visualization of element groups and hash codes.}
\label{fig:vis}
\end{figure*}

\subsection{Results}

We next present the results from our computational study on synthetic data. Let $Z$ denote the learned hash code; for elements $i \in \mathcal{S}_0$, $\*z_i$ is obtained using one of the algorithms presented in Section \ref{sec:learn-opt-hash}; for elements $i \not\in \mathcal{S}_0$, $\*z_i$ is obtained using machine learning, as per Section \ref{sec:similarity-based-estimation}. Then, the metrics that we consider are the estimation error $\sum_{i \in [n]} \sum_{j \in [b]} z_{ij} \left| f_i^0 - \frac{\sum_{k \in [n]} z_{kj} f^0_k}{\sum_{k \in [n]} z_{kj}} \right|$, the similarity error $\sum_{i \in [n]} \sum_{j \in [b]} z_{ij} \sum_{k \in [n]} z_{kj} \| \*x_i - \*x_k \|^2$, and the overall error, i.e., the convex combination of the above two error terms, weighted by $\lambda$ and $1-\lambda$, respectively, which is exactly the objective function that we use in the proposed formulation. We separately study the two error terms to shed light on the trade-off that the proposed approach is faced with. Moreover, we distinguish between the error on elements which appeared in the prefix (and hence their estimate is extracted from the learned hashing scheme) and the error on unseen elements which did not appear in the prefix (and hence their estimate is inferred using machine learning) to examine the individual performance of each component of the proposed approach. We also measure the running time (in seconds) of the algorithms (note that the running time includes the time to learn both the hashing scheme and the classifier).

\hspace{0.25pt}

\textbf{Experiment 1: Impact of hyperparameter $\lambda$.} In this experiment, we study the impact of the hyperparameter $\lambda$ on the learned hashing scheme. We set $G=6$ and run three different versions of \oh for varying $\lambda$: one that uses \milp\ to learn the hashing scheme, one uses \bcd, and one uses \dprog. We record the estimation, similarity, and overall error on the prefix, as well as the running time of each algorithm. To examine the degree of sub-optimality of \bcd, we present the actual values of the error terms that constitute the objective function (estimation, similarity, and overall error), i.e., we do not convert them in a per element/per pair of elements scale, which would be more interpretable. The results are presented in Figure \ref{fig:exper1}. The key takeaways from this experiment are as follows:
\begin{itemize}
    \item[-] \milp\ achieves the smallest overall error at the cost of increased running times. Its edge over the heuristic \bcd\ approach can be verified in terms of the estimation error, as it almost always improves over the solution obtained by \bcd.
    \item[-] The solutions obtained by \bcd\ are of high quality; the improvement achieved by applying the exact \milp\ approach is often negligible. For small problem sizes, the runtime of \bcd\ is less than a second.
    \item[-] As expected, \dprog\ achieves the smallest estimation error, since it optimizes only for the estimation error independently of the value of $\lambda$. In terms of the similarity and the overall, the performance of \dprog\ is notably worse, whereas its running time is less than a second.
\end{itemize}
Note that, in the $\lambda=1$ case, all three methods are able to find comparable near-optimal solutions. The small deviation is due to suboptimality tolerances of the algorithms used.

\begin{figure*}[!ht]
\centering
\subfloat[Estimation error on $\mathcal{S}_0$. ]{\includegraphics[width=0.48\columnwidth]{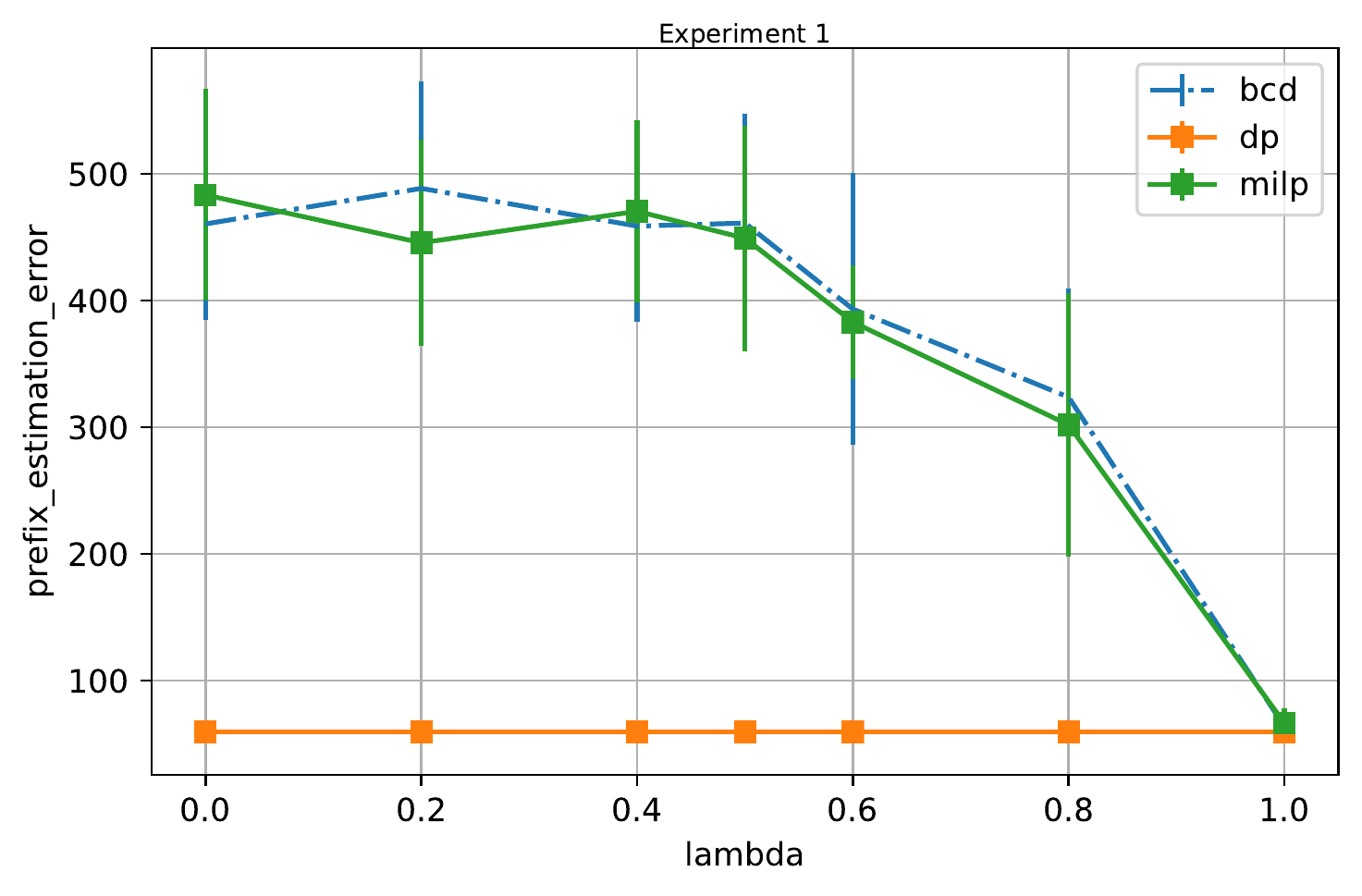}%
\label{fig:exper1_estimation}}
\hfil
\subfloat[Similarity error on $\mathcal{S}_0$. ]{\includegraphics[width=0.48\columnwidth]{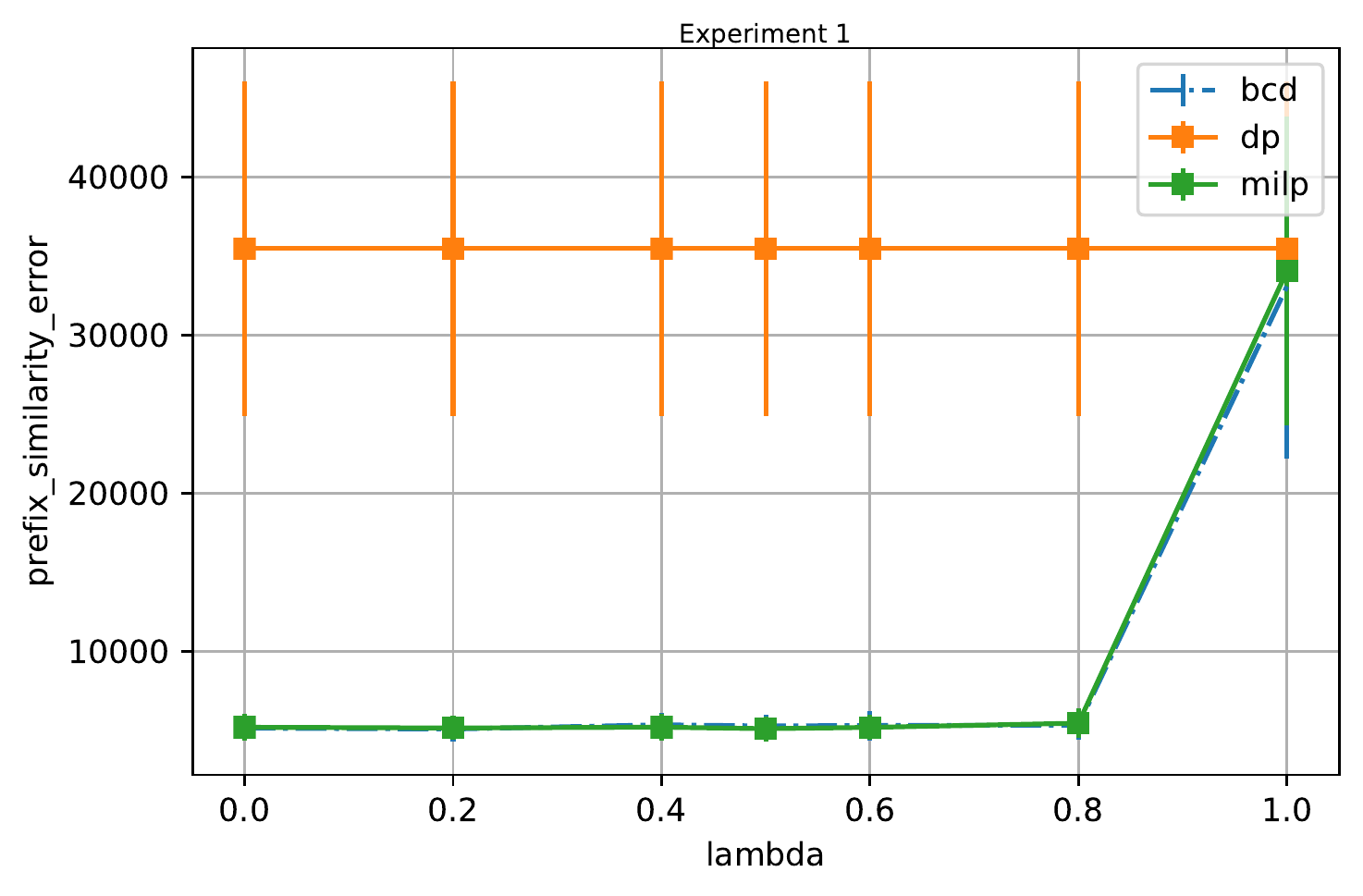}%
\label{fig:exper1_similarity}}
\hfil
\subfloat[Overall error (objective function value) on $\mathcal{S}_0$.]{\includegraphics[width=0.48\columnwidth]{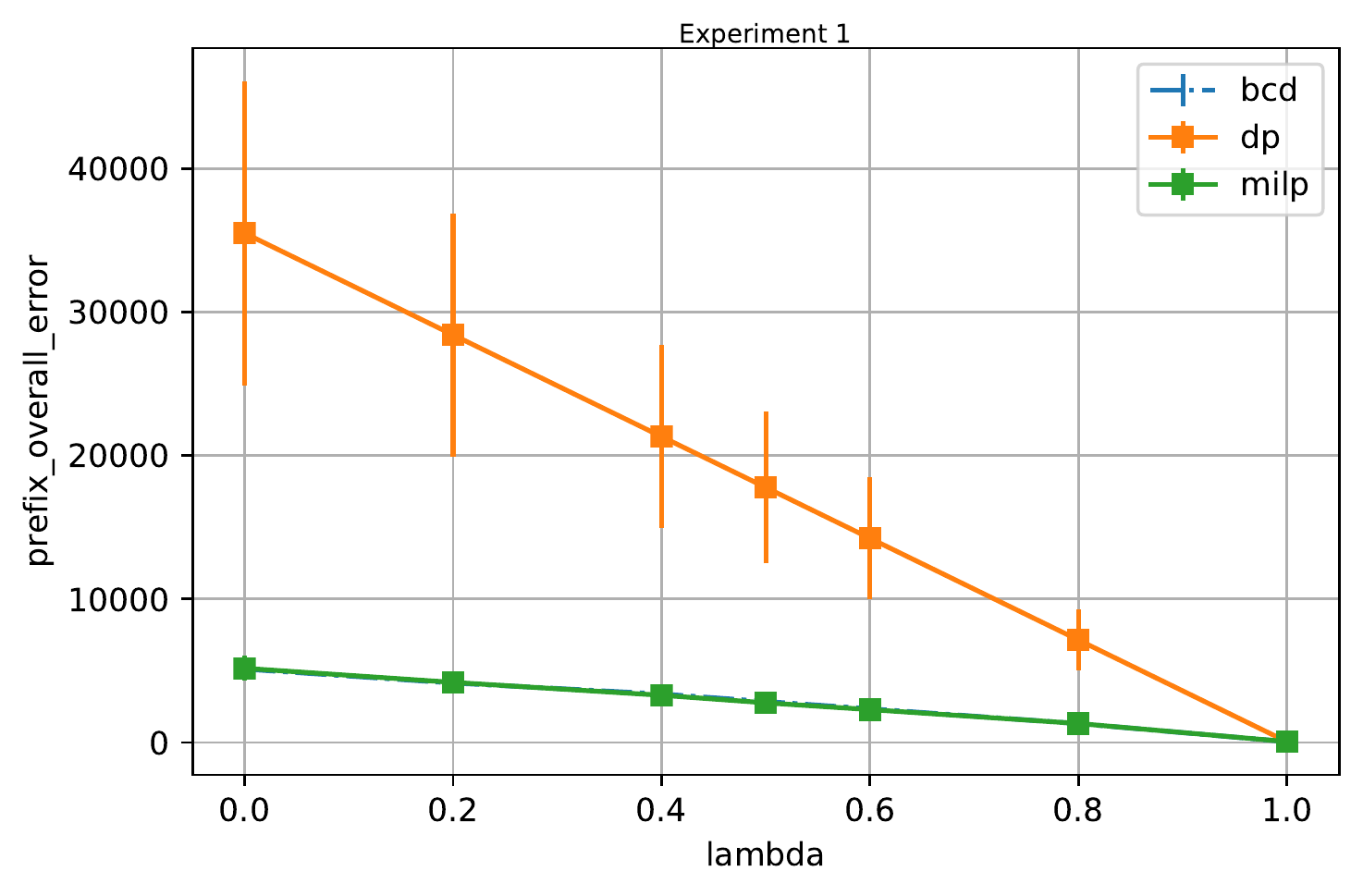}%
\label{fig:exper1_overall}}
\hfil
\subfloat[Elapsed time (in sec). ]{\includegraphics[width=0.48\columnwidth]{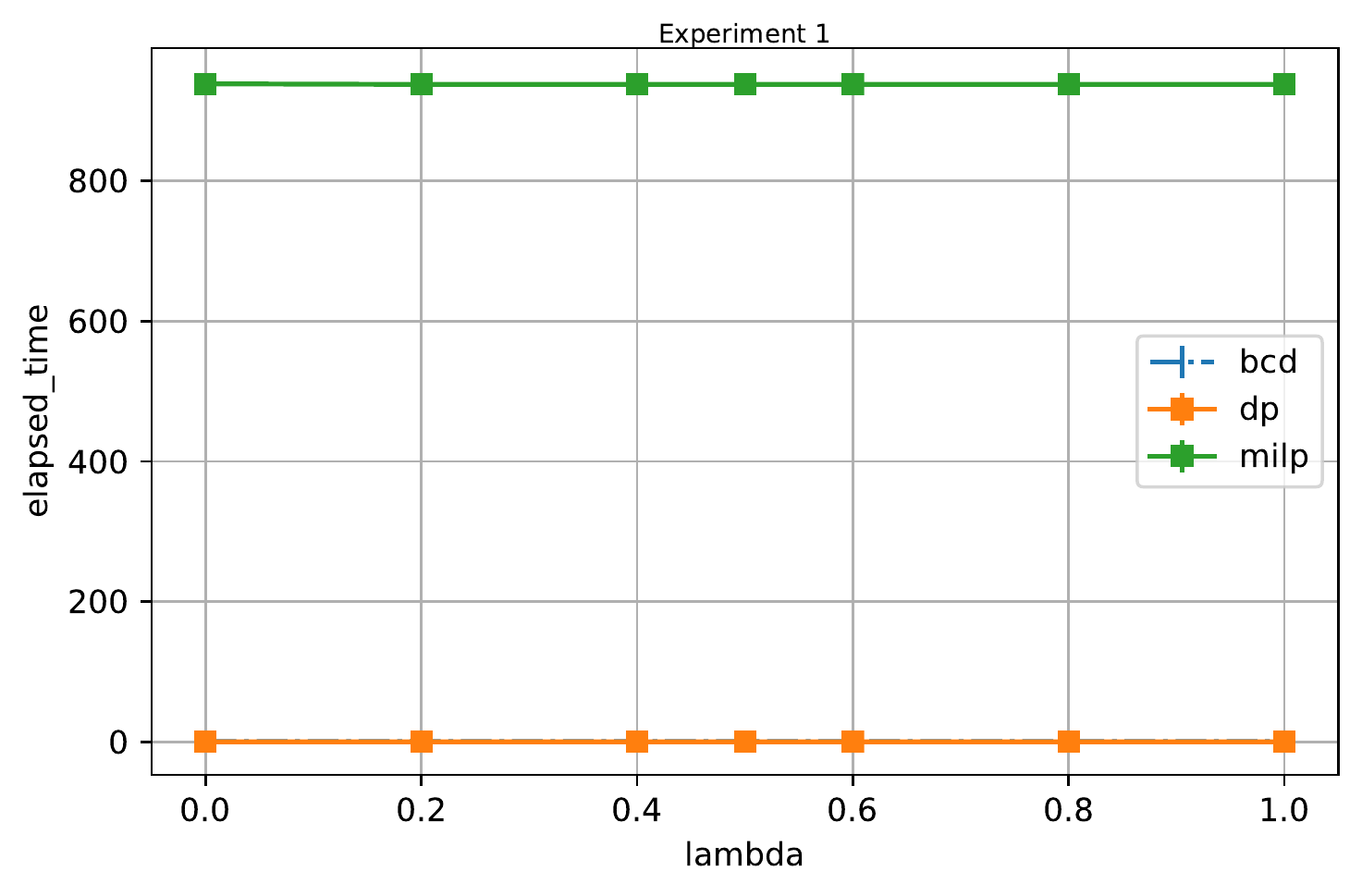}%
\label{fig:exper1_time}}
\caption{Impact of hyperparameter $\lambda$ for $G=6$.}
\label{fig:exper1}
\end{figure*}


\hspace{0.25pt}

\textbf{Experiment 2: Comparison between \bcd\ and \dprog\ in the $\lambda=1$ case.} In this experiment, we focus on the $\lambda=1$ case and compare, for increasing values of $G$, \bcd\ with \dprog; in this case, the latter is guaranteed to find the optimal hashing scheme. We again record the estimation, similarity, and overall error on the prefix, as well as the running time of each algorithm. In this and in subsequent experiment, we convert the errors in a per element/per pair of elements scale. The results are presented in Figure \ref{fig:exper2}. We observe that, for problems with $G\leq10$, \bcd\ computes near-optimal solutions fast; however, as $G$ further increases, the performance of \bcd\ deteriorates.

\begin{figure*}[!ht]
\centering
\subfloat[Estimation error on $\mathcal{S}_0$. ]{\includegraphics[width=0.48\columnwidth]{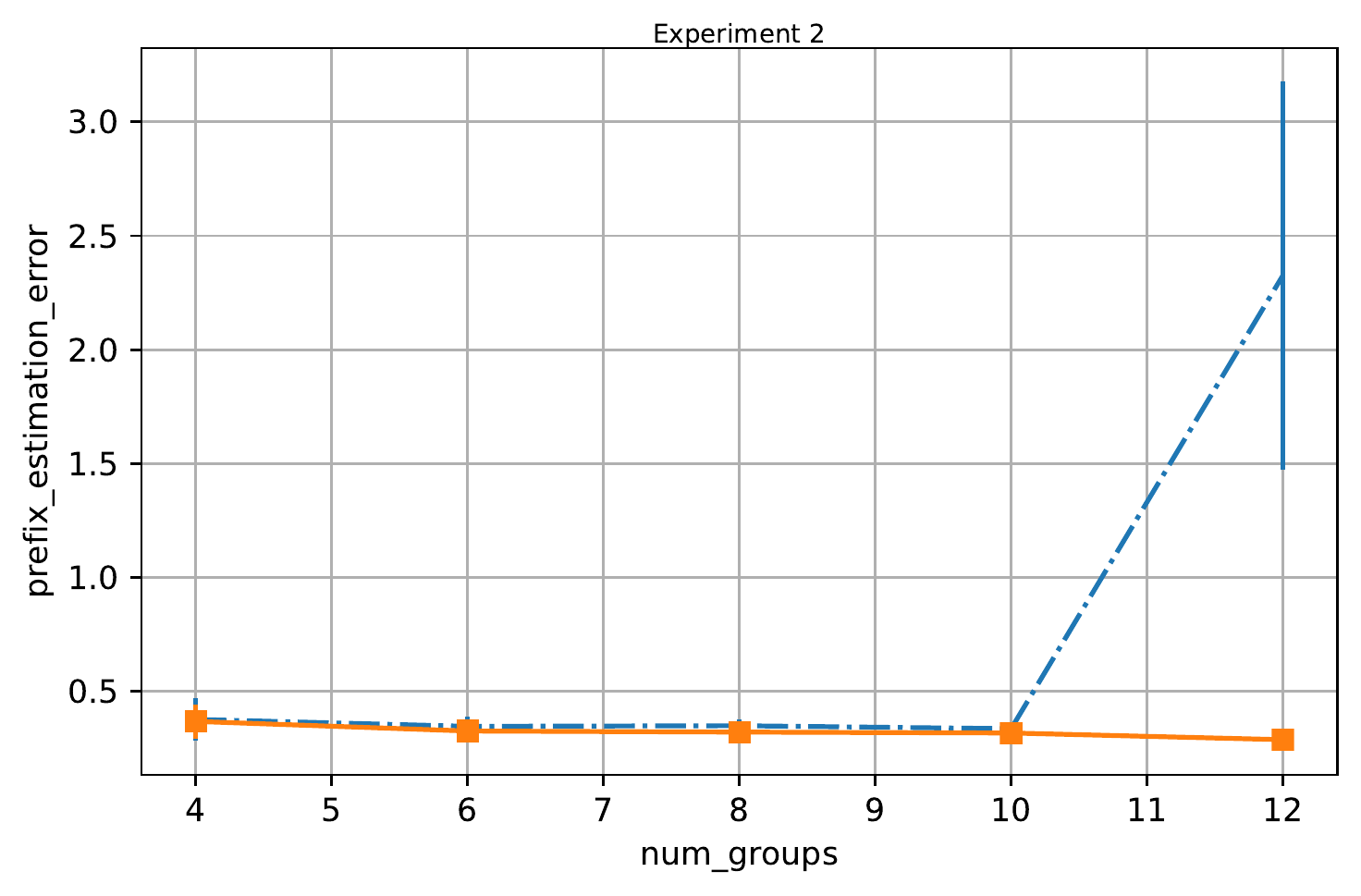}%
\label{fig:exper2_estimation}}
\hfil
\subfloat[Similarity error on $\mathcal{S}_0$. ]{\includegraphics[width=0.48\columnwidth]{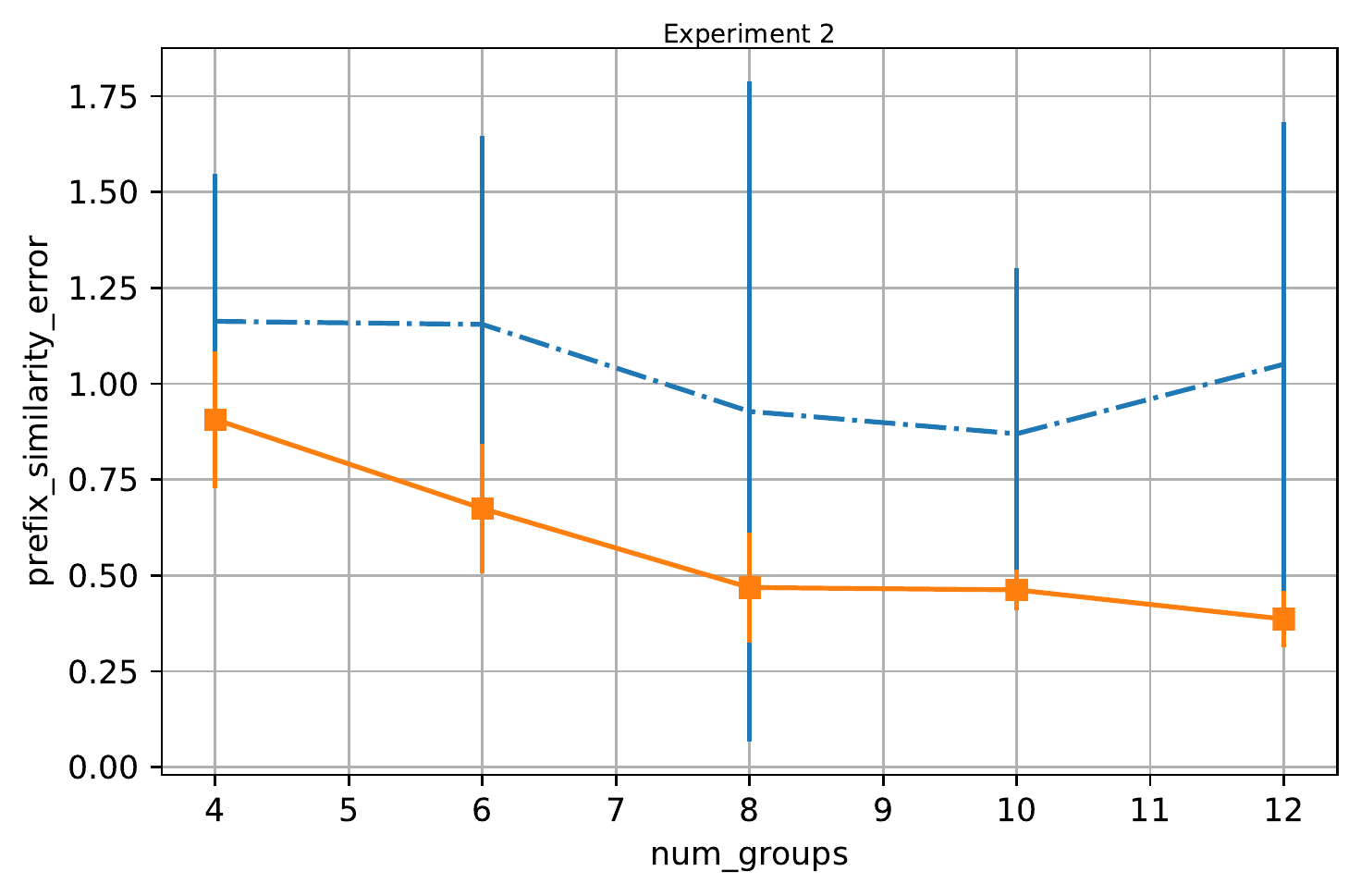}%
\label{fig:exper2_similarity}}
\hfil
\subfloat[Overall error (objective function value) on $\mathcal{S}_0$. ]{\includegraphics[width=0.48\columnwidth]{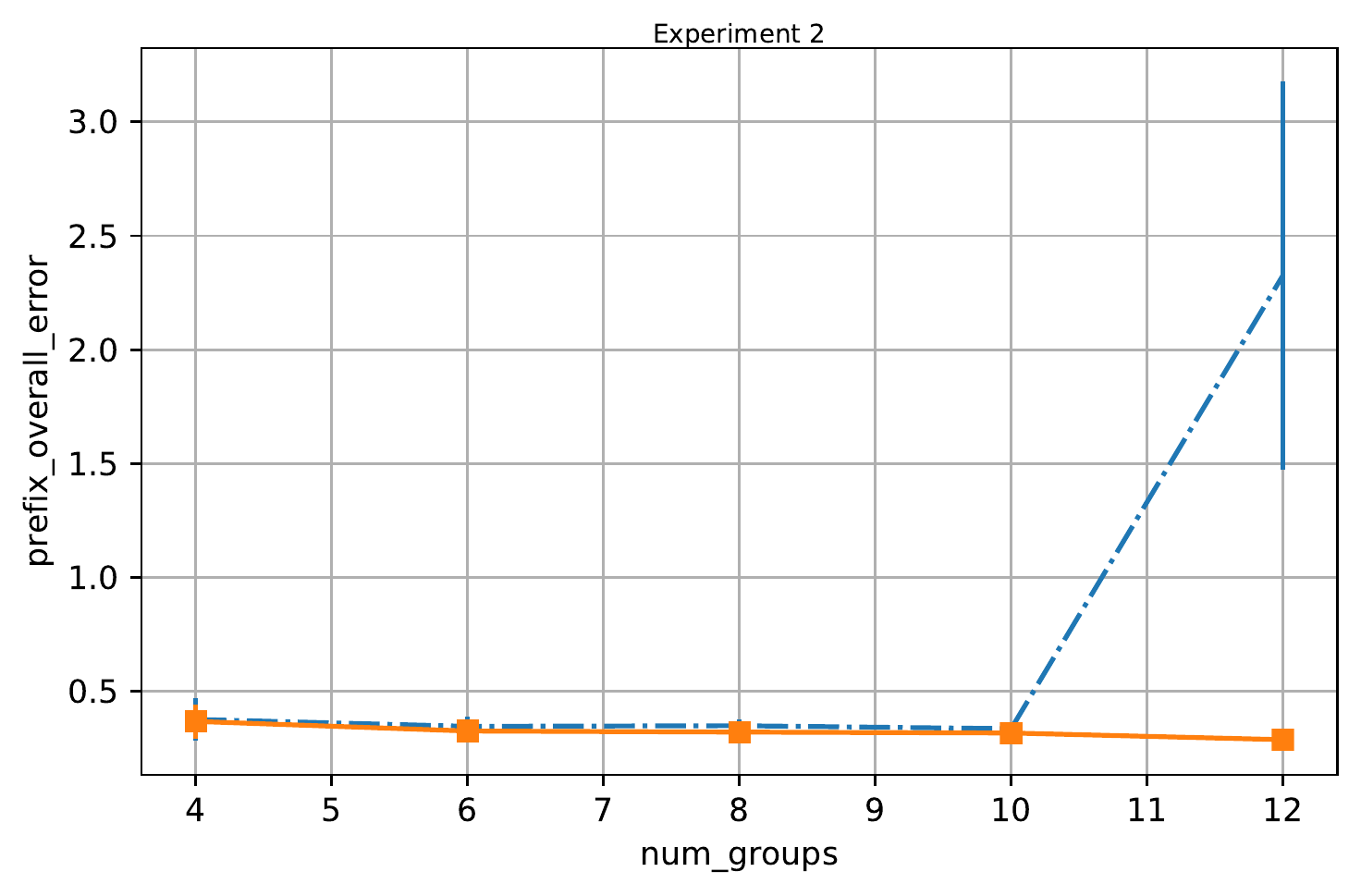}%
\label{fig:exper2_overall}}
\hfil
\subfloat[Elapsed time (in sec). ]{\includegraphics[width=0.48\columnwidth]{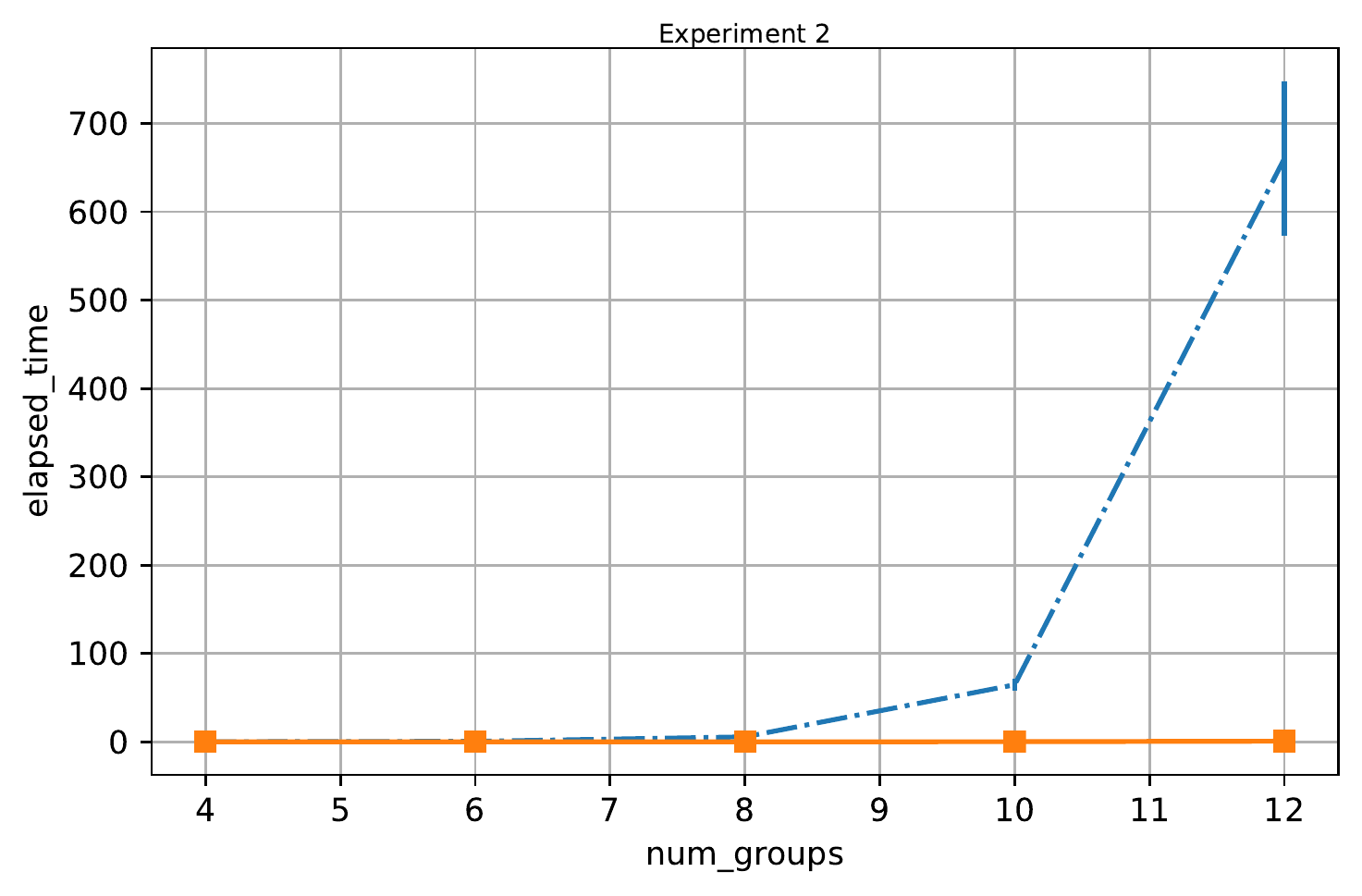}%
\label{fig:exper2_time}}
\caption{Comparison between \dprog\ and \bcd\ for $\lambda=1$.}
\label{fig:exper2}
\end{figure*}

\hspace{0.25pt}

\textbf{Experiment 3: \bcd\ from multiple starting points in the general $\lambda$ case.} In this experiment, we set $\lambda=0.5$ and run \bcd\ multiple times from different starting points and for increasing values of $G$ to examine the stability of the solutions obtained. We again record the estimation, similarity, and overall error on the prefix, as well as the running time of each algorithm. The results, presented in Figure \ref{fig:exper3}, indicate that \bcd\ is robust to the (random) initialization of the algorithm and computes stable solutions.

\begin{figure*}[!ht]
\centering
\subfloat[Estimation error on $\mathcal{S}_0$. ]{\includegraphics[width=0.48\columnwidth]{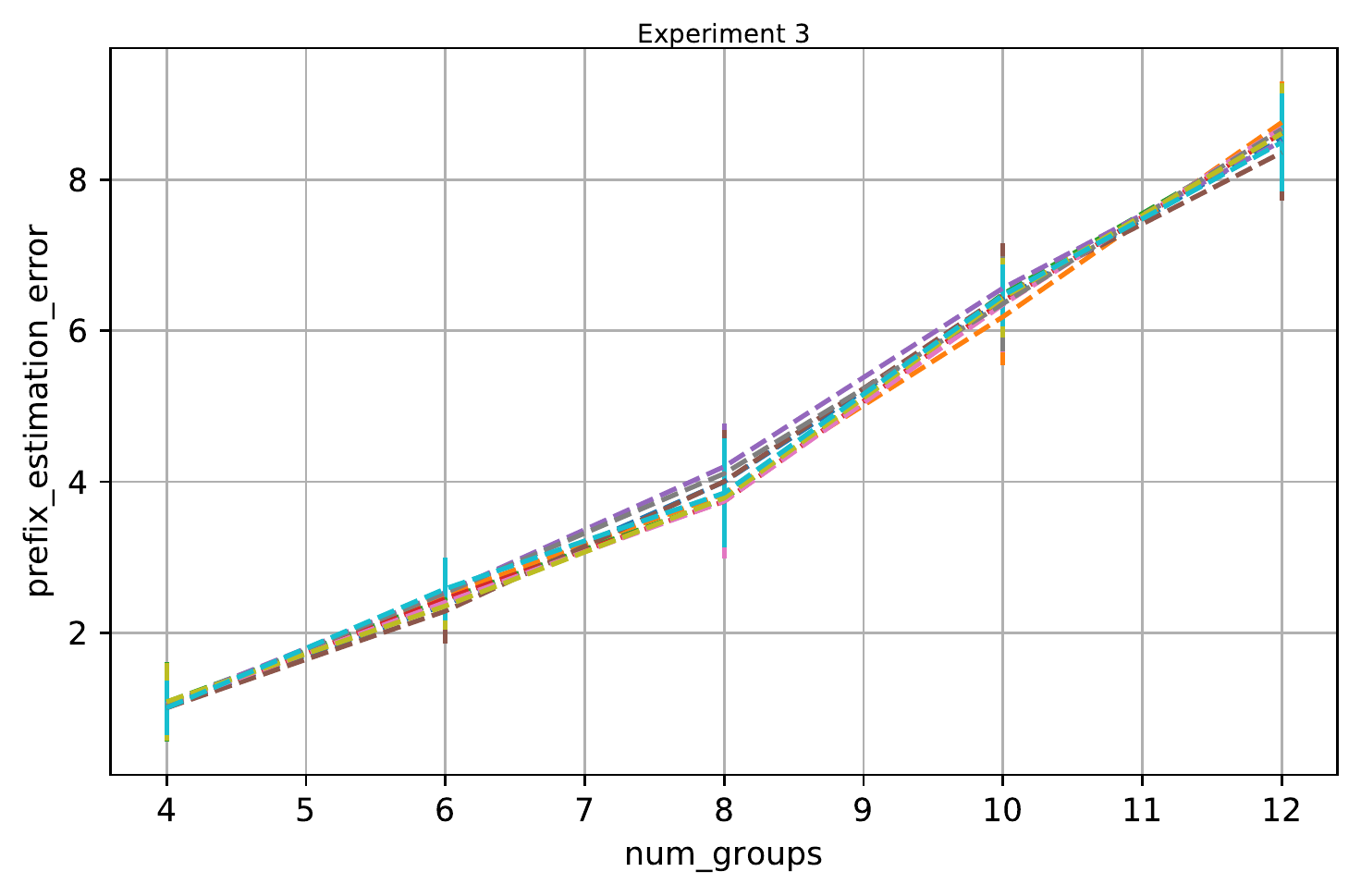}%
\label{fig:exper3_estimation}}
\hfil
\subfloat[Similarity error on $\mathcal{S}_0$. ]{\includegraphics[width=0.48\columnwidth]{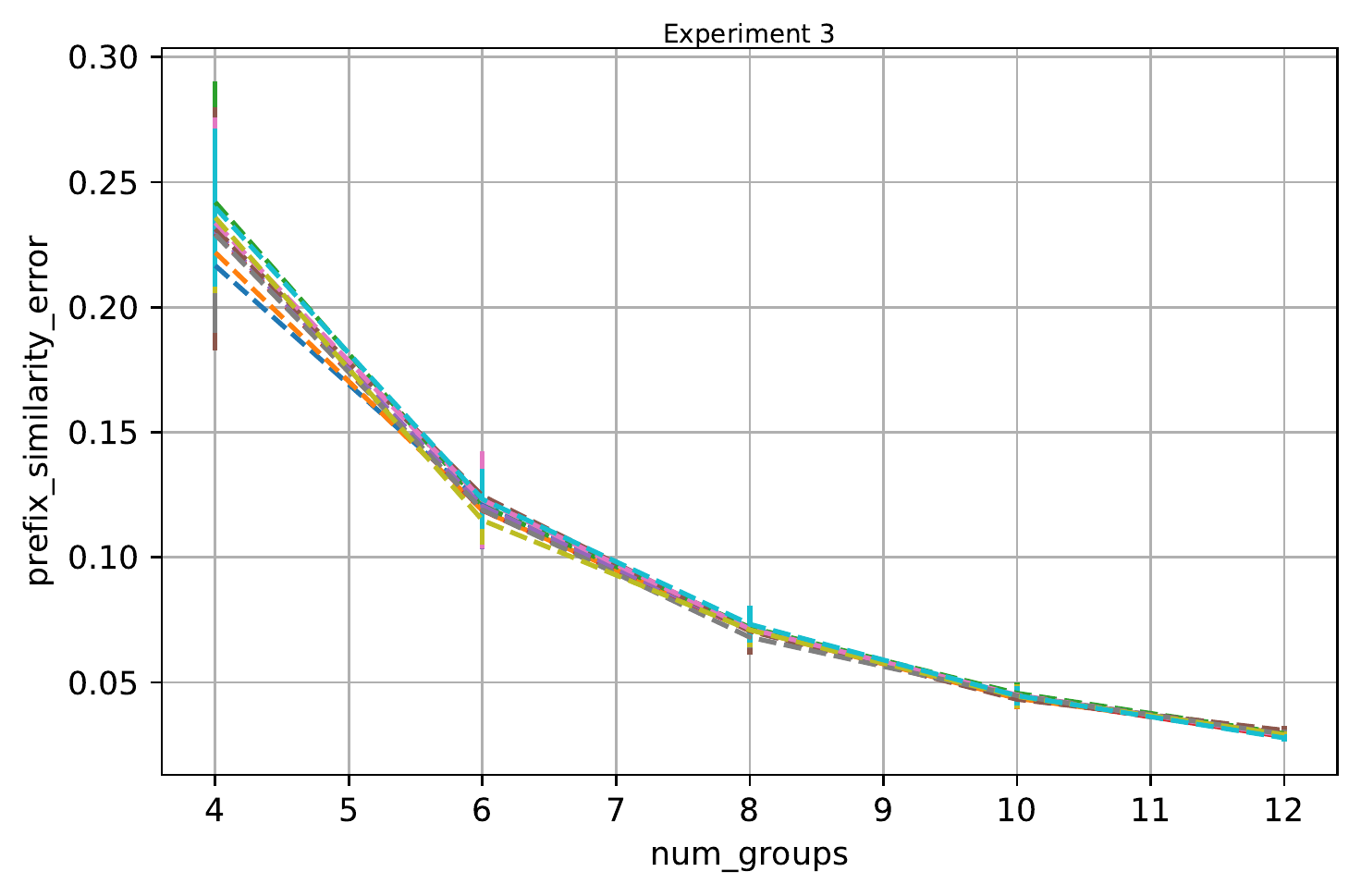}%
\label{fig:exper3_similarity}}
\hfil
\subfloat[Overall error (objective function value) on $\mathcal{S}_0$. ]{\includegraphics[width=0.48\columnwidth]{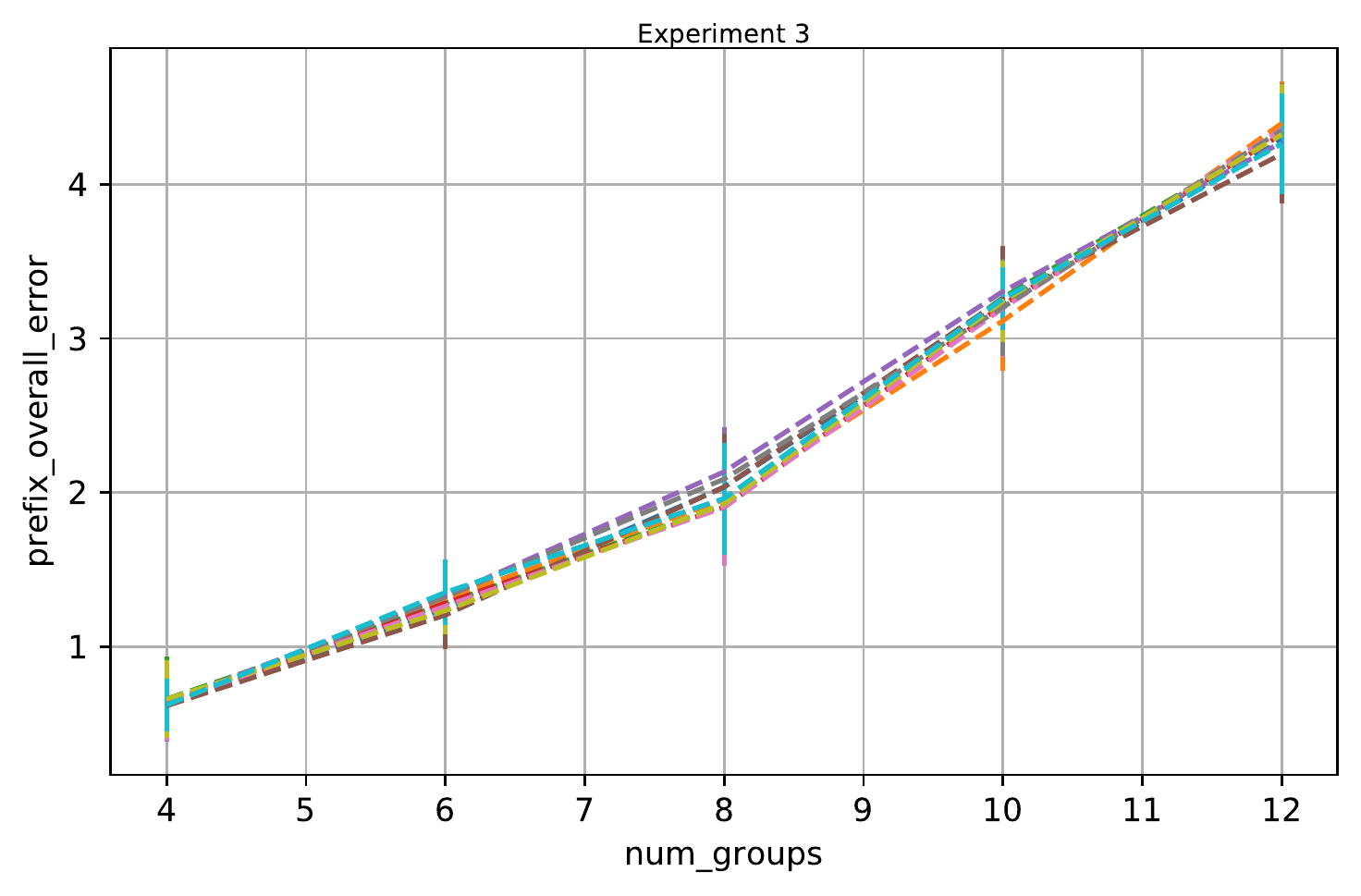}%
\label{fig:exper3_overall}}
\hfil
\subfloat[Elapsed time (in sec). ]{\includegraphics[width=0.48\columnwidth]{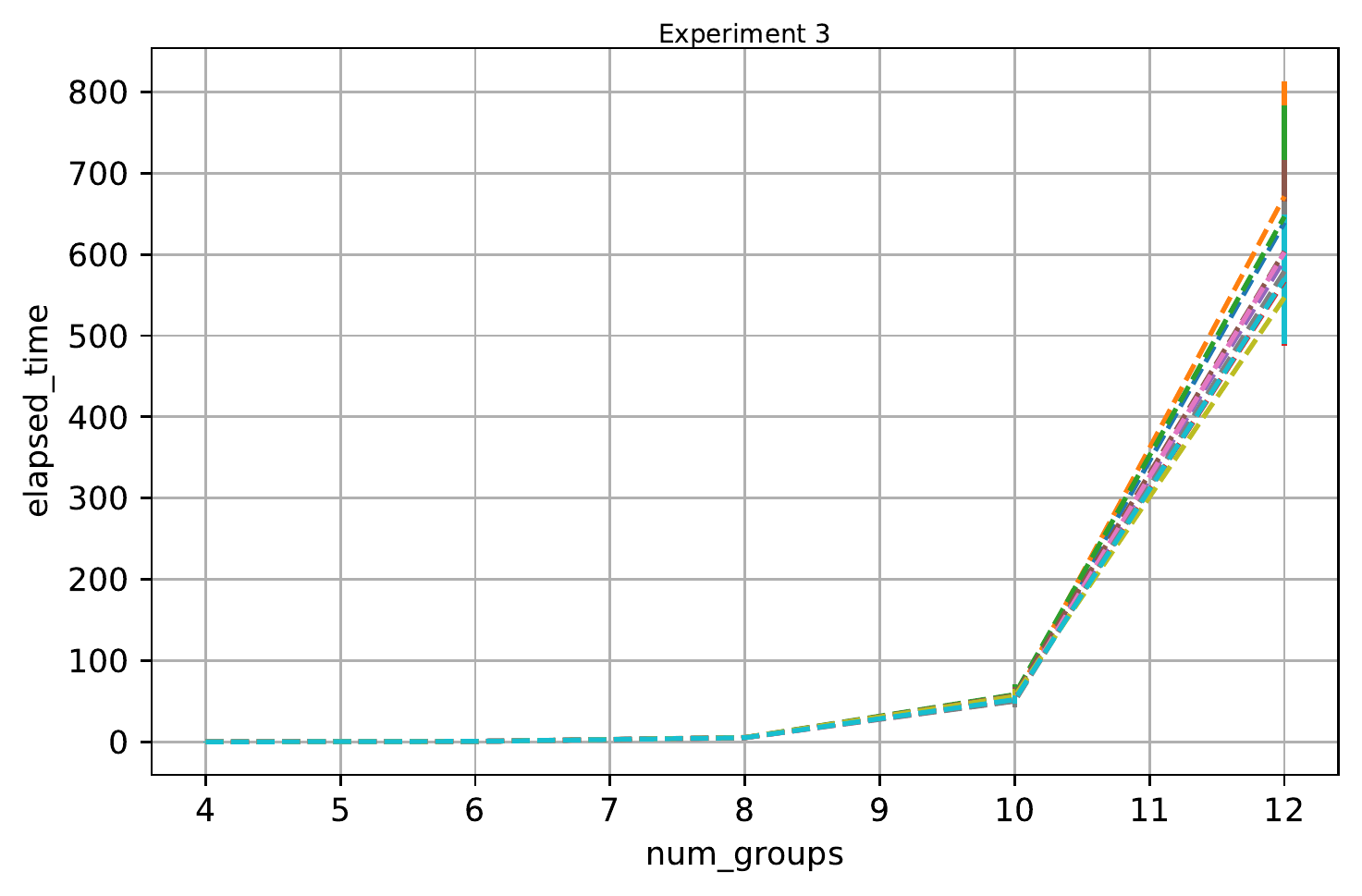}%
\label{fig:exper3_time}}
\caption{Comparison between \bcd\ from multiple starting points for $\lambda=0.5$.}
\label{fig:exper3}
\end{figure*}

\hspace{0.25pt}

\textbf{Experiment 4: Impact of the fraction of elements seen in the prefix.}  In this experiment, we set $G=10$ and vary the value of $g_0,$ which controls the fraction of elements that appear in the prefix. We explore two approaches for learning the hashing scheme: first, we set $\lambda=0.5$ and run \bcd; then, we run \dprog\ (which implies $\lambda=1$). We now record the estimation and similarity error both on the prefix $\mathcal{S}_0$ and on elements that did not appear in $\mathcal{S}_0$ but did appear within $|\mathcal{S}| = 10|\mathcal{S}_0|$ arrivals after $\mathcal{S}_0$. Figure \ref{fig:exper4} suggests that observing more elements in the prefix results in a decrease of the estimation error on both seen and unseen elements at the cost of an increased similarity error.

\begin{figure*}[!ht]
\centering
\subfloat[Estimation error on $\mathcal{S}_0$. ]{\includegraphics[width=0.48\columnwidth]{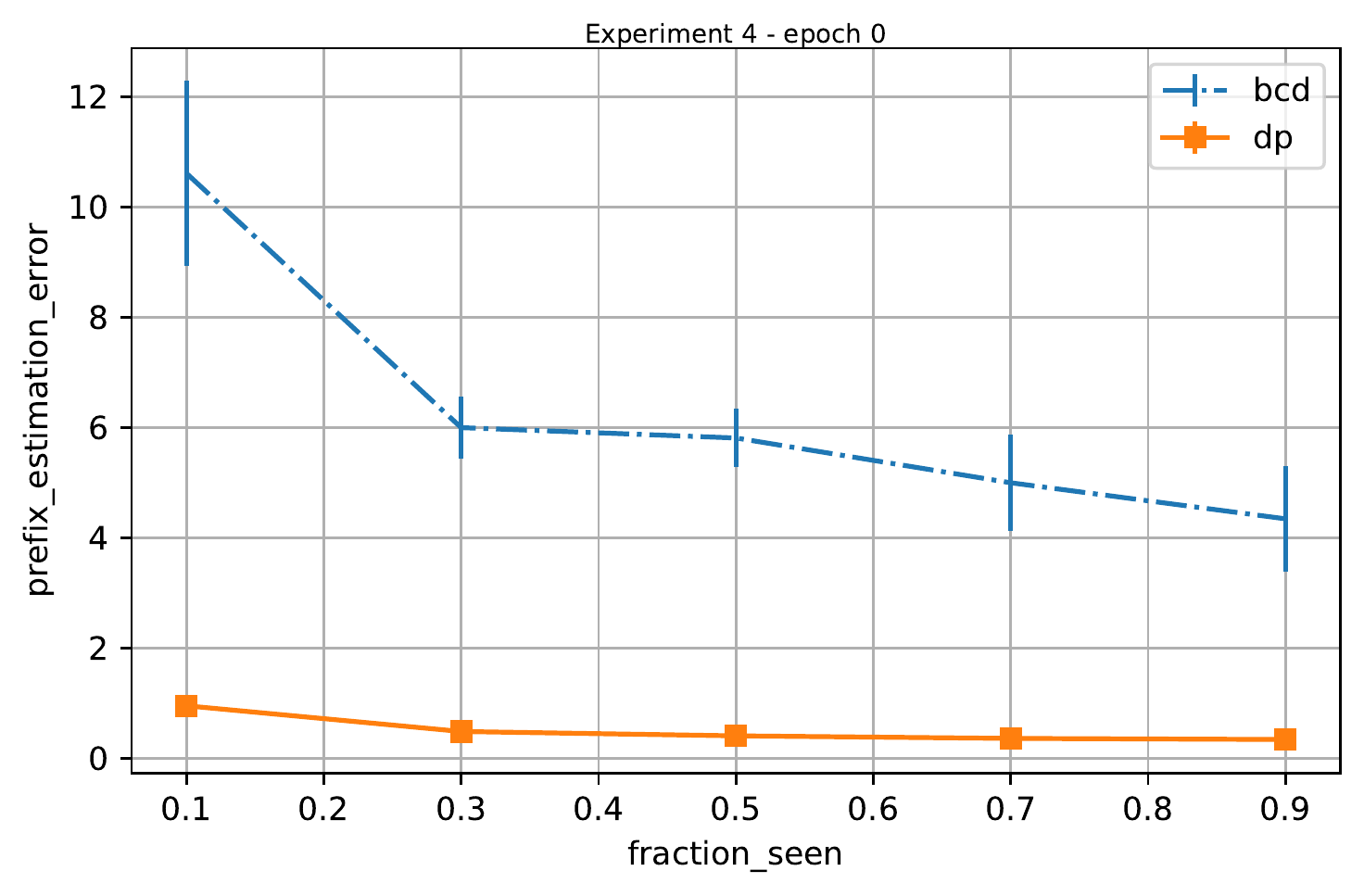}%
\label{fig:exper4_estimation}}
\hfil
\subfloat[Similarity error on $\mathcal{S}_0$. ]{\includegraphics[width=0.48\columnwidth]{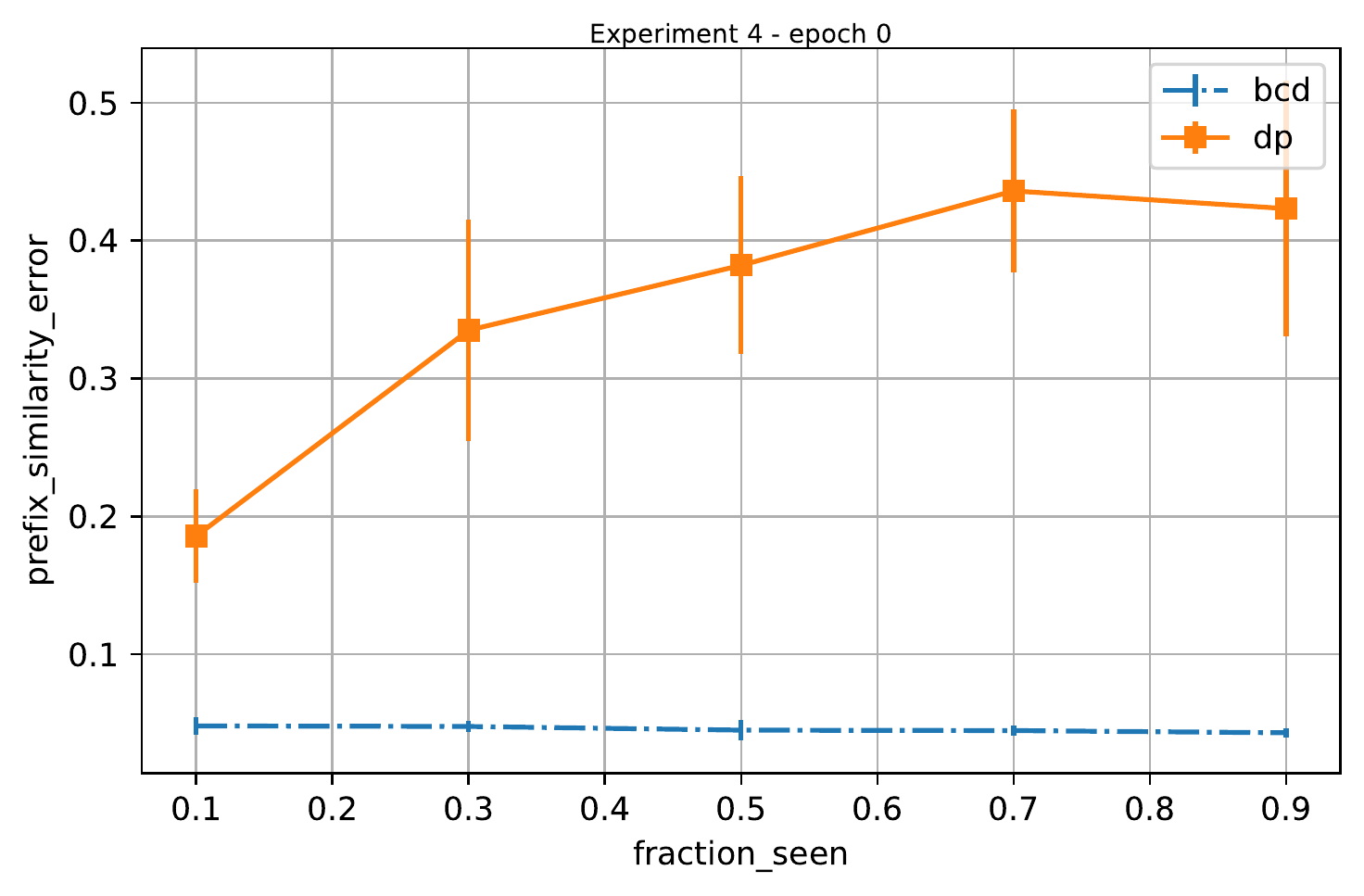}%
\label{fig:exper4_similarity}}
\hfil
\subfloat[Estimation error on elements $u \not\in \mathcal{S}_0$ after $|\mathcal{S}| = 10|\mathcal{S}_0|$ arrivals. ]{\includegraphics[width=0.48\columnwidth]{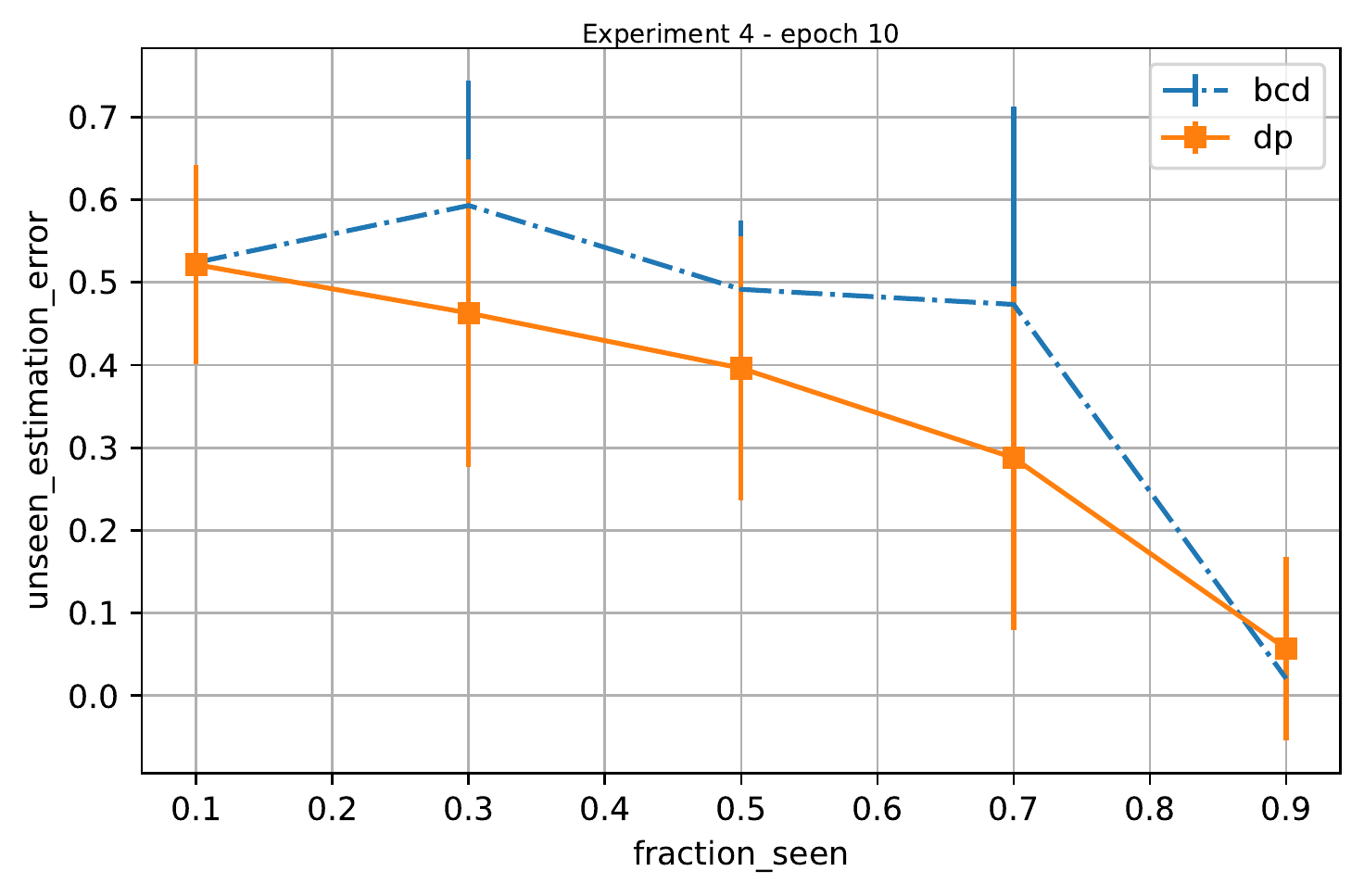}%
\label{fig:exper4_unseen_estimation}}
\hfil
\subfloat[Similarity error on elements $u \not\in \mathcal{S}_0$ after $|\mathcal{S}| = 10|\mathcal{S}_0|$ arrivals. ]{\includegraphics[width=0.48\columnwidth]{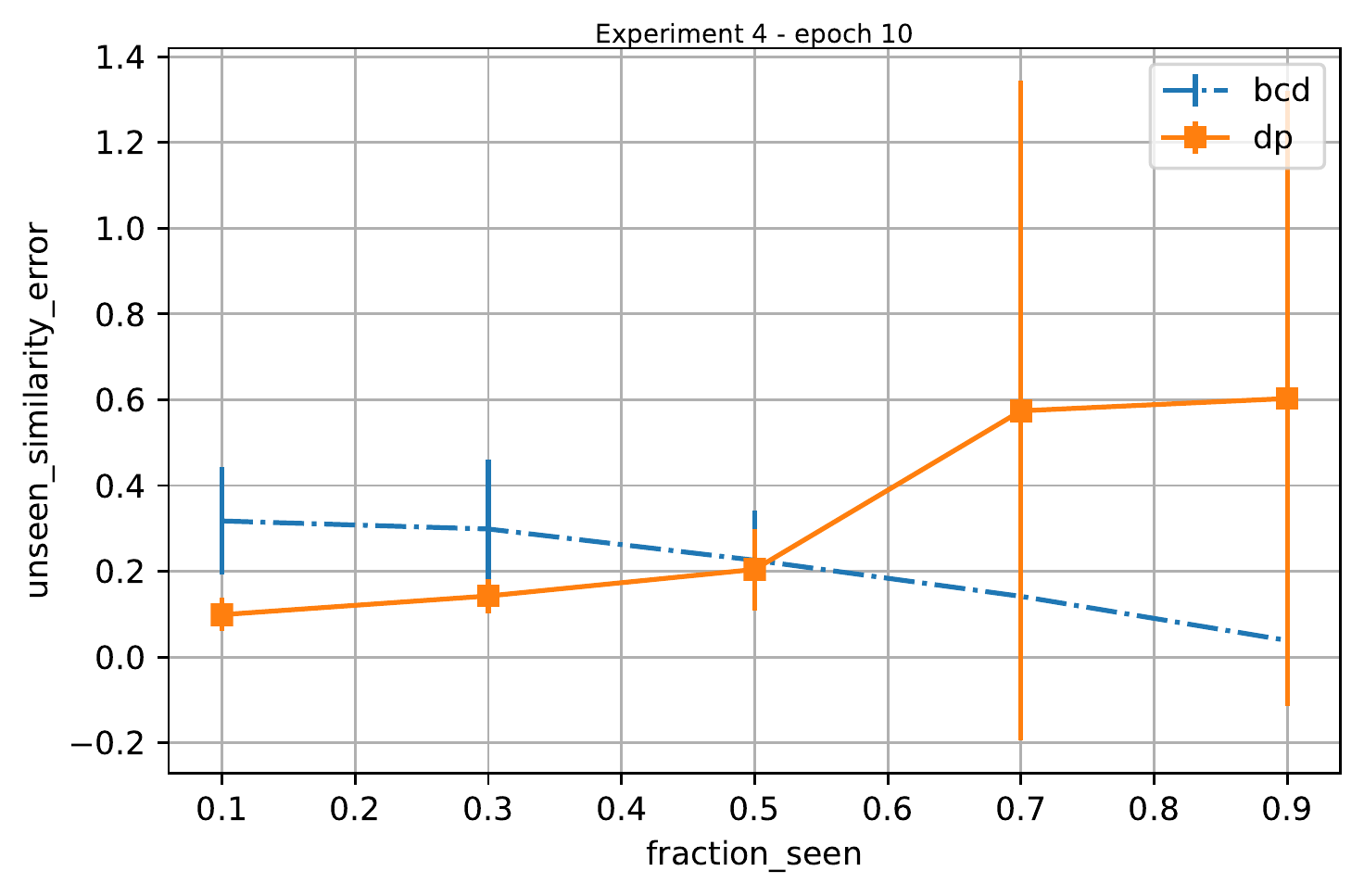}%
\label{fig:exper4_unseen_similarity}}
\caption{Impact of fraction of seen elements in the prefix ($g_0$) for $G=10$.}
\label{fig:exper4}
\end{figure*}

\hspace{0.25pt}

\textbf{Experiment 5: Comparison between classification methods.} In this experiment, we set $g_0=0.33$ and $\lambda=0.5$, vary the value of $G$, and explore the impact of using different types of classifiers (\logreg, \cart, \rf) as part of \oh. We record the estimation, similarity, and overall error on elements that did not appear in $\mathcal{S}_0$ but did appear within $|\mathcal{S}| = 10|\mathcal{S}_0|$ arrivals after $\mathcal{S}_0$. We also report the training time for each method. In Figure \ref{fig:exper5}, we see that there is indeed merit in using non-linear classifiers. We remark, however, that the results heavily depend on the data generating process.

\begin{figure*}[!ht]
\centering
\subfloat[Estimation error on elements $u \not\in \mathcal{S}_0$ after $|\mathcal{S}| = 10|\mathcal{S}_0|$ arrivals. ]{\includegraphics[width=0.48\columnwidth]{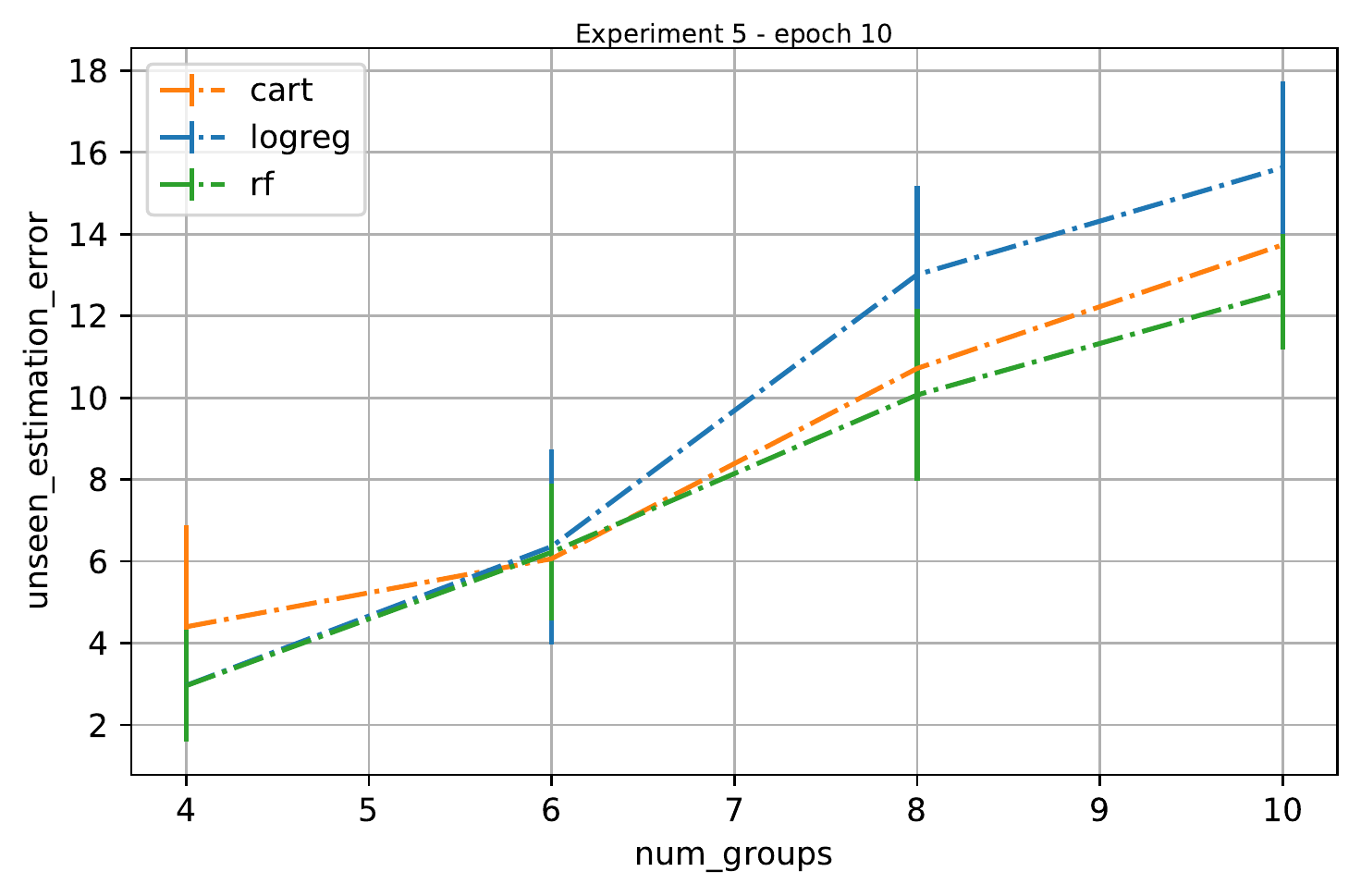}%
\label{fig:exper5_estimation}}
\hfil
\subfloat[Similarity error on elements $u \not\in \mathcal{S}_0$ after $|\mathcal{S}| = 10|\mathcal{S}_0|$ arrivals. ]{\includegraphics[width=0.48\columnwidth]{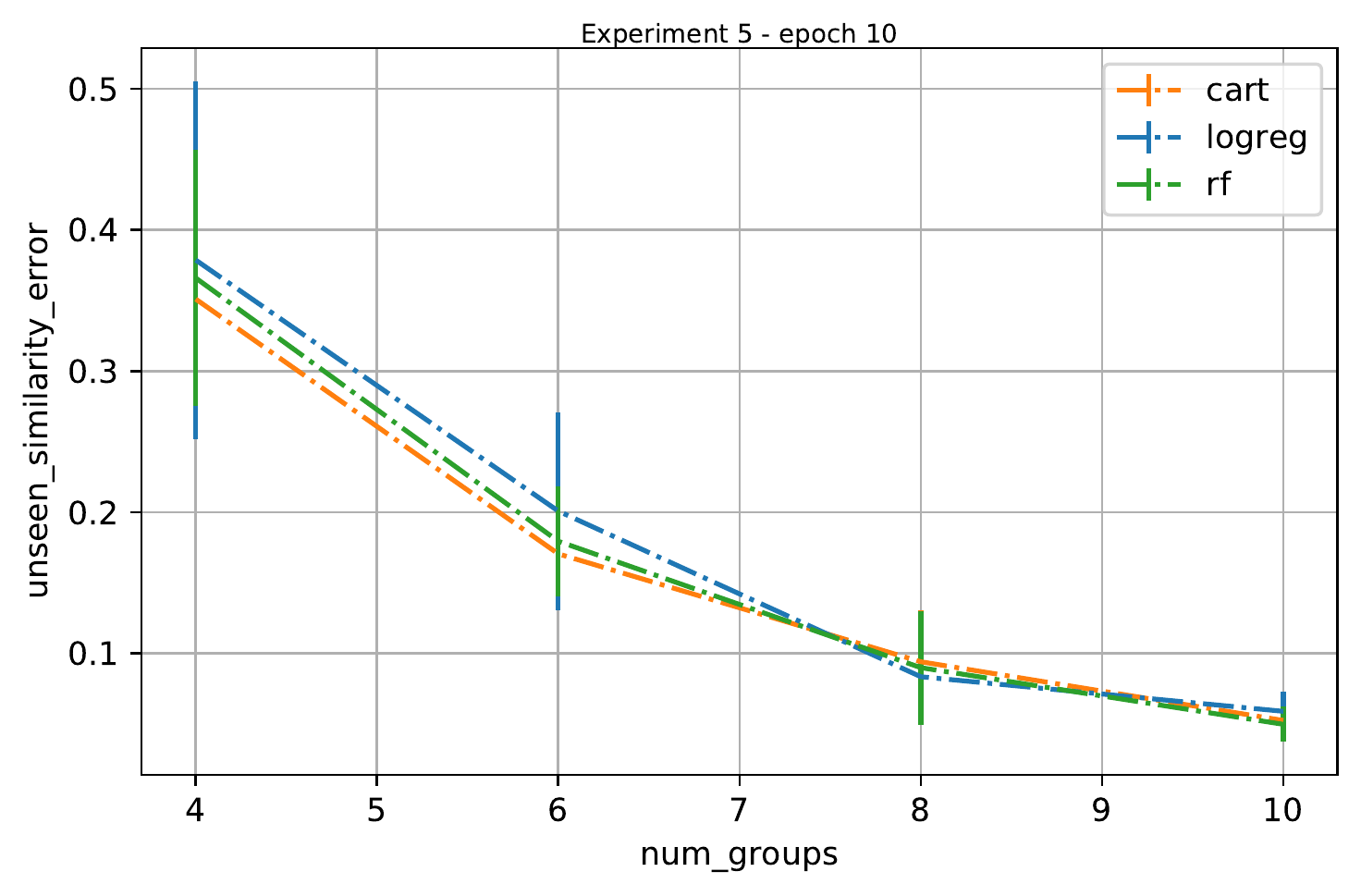}%
\label{fig:exper5_similarity}}
\hfil
\subfloat[Overall error (objective function value) on elements $u \not\in \mathcal{S}_0$ after $|\mathcal{S}| = 10|\mathcal{S}_0|$ arrivals. ]{\includegraphics[width=0.48\columnwidth]{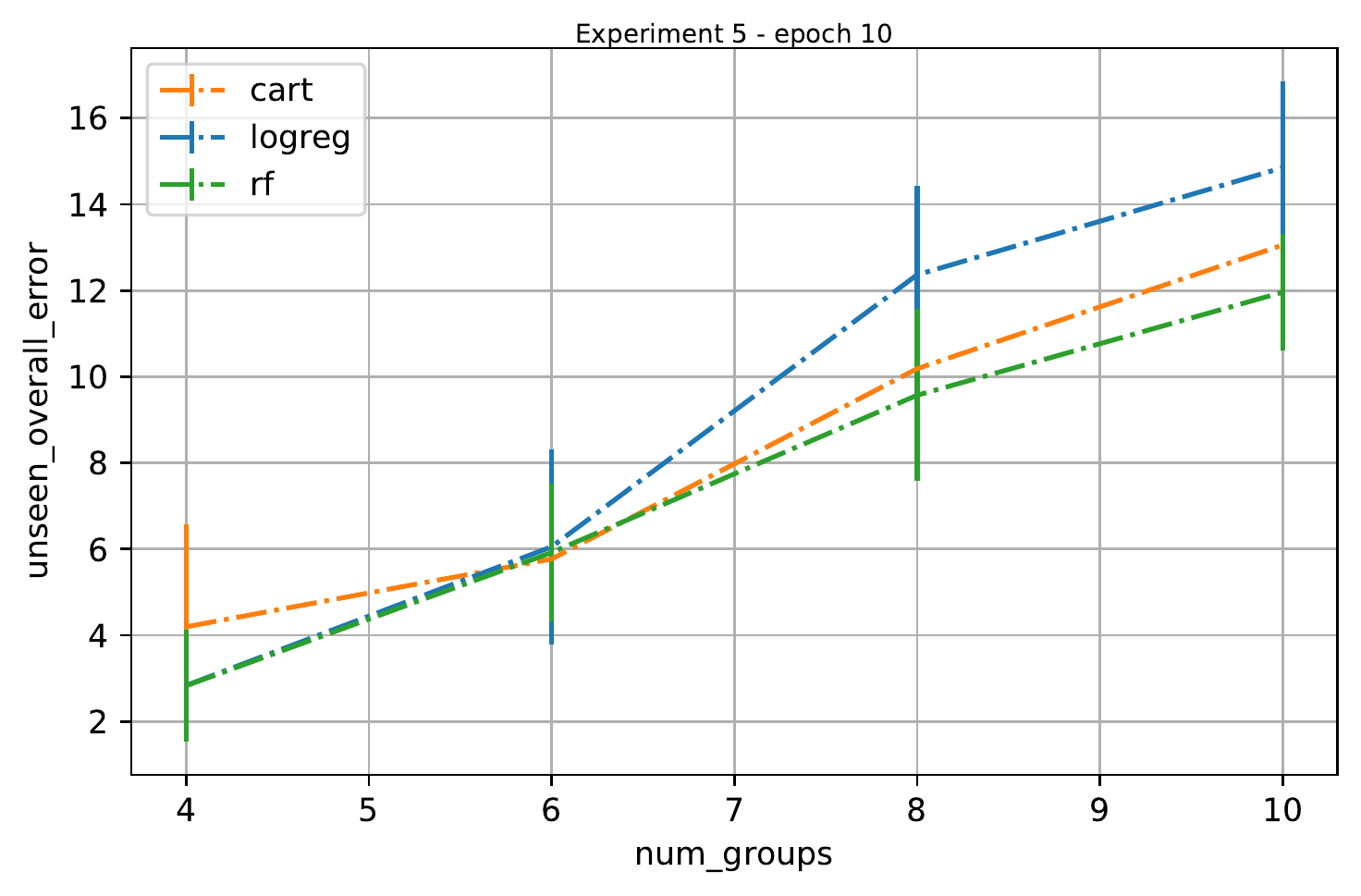}%
\label{fig:exper5_overall}}
\hfil
\subfloat[Elapsed time (in sec). ]{\includegraphics[width=0.48\columnwidth]{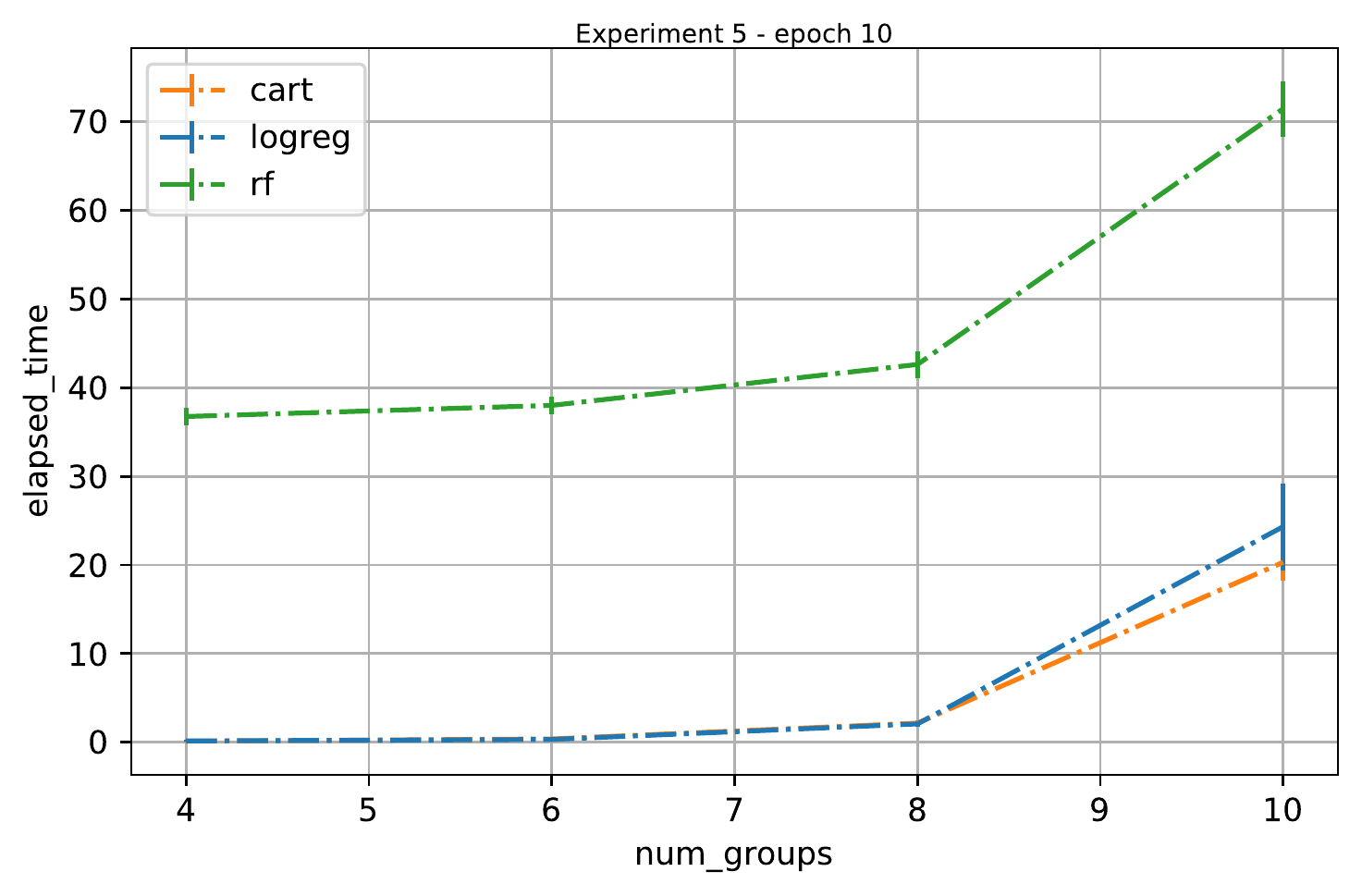}%
\label{fig:exper5_time}}
\caption{Comparison between classification methods.}
\label{fig:exper5}
\end{figure*}


}

\section{Experiments on Real-World Data: Search Query Estimation} \label{sec:experiments}
In this section, we empirically evaluate the proposed approach on real-world search query data. The task of search query frequency estimation seems particularly suited for the proposed learning-based approach, given that popular search queries tend to appear consistently across multiple days.

\subsection{Dataset}
In the lines of \cite{hsu2018learningbased}, we use the AOL query log dataset, which consists of 21 million search queries (with 3.8 million unique ones) collected from 650 thousand anonymized users over 90 days in 2006. Each query is a search phrase in free text; for example, the $1^{st}$ most common query is ``google'' and appears 251,463 times over the entire 90-day period, the $10^{th}$ is ``www.yahoo.com'' and its frequency is 37,436, the $100^{th}$ is ``mys'' and its frequency is 5,237, the $1000^{th}$ is ``sharon stone'' and its frequency is 926, the $10000^{th}$ is ``online casino'' and its frequency is 146, and so forth. As shown in \cite{hsu2018learningbased}, the distribution of search query frequency indeed follows the Zipfian law and hence the setting seems ideal for their proposed algorithm (LCMS).

\subsection{Baselines}
As baselines, we use \cm\ and \hh. For each method, we maintain multiple versions corresponding to different values of the method's hyperparameters and report the best performing version. More specifically, for fixed sketch size (i.e., total number of buckets $b$), we report the best performing for \cm's depth from the set $d \in \{1,2,4,6\}$ and for \hh's depth $d \in \{1,2,4,6\}$ and number of heavy-hitter buckets $b_{heavy} \in \{10,10^2,10^3,10^4\}$ (provided that $b_{heavy}$ fits within the available memory, i.e., $b_{heavy} \leq \nicefrac{b}{2}$). Additionally, we assume that \hh\ has access to an ideal heavy-hitter oracle, i.e., the IDs of the heavy-hitters in the test set (over the entire 90-day period) are known. Therefore, we compare the proposed method with the ideal version of the method proposed in \cite{hsu2018learningbased}, which was in fact shown to significantly outperform any realistically implementable version of \hh\ that relied upon non-ideal heavy-hitter oracles (e.g., recurrent neural network classifier).

\subsection{Remarks on the Learned Hashing Scheme}
As far as \oh\ is concerned, we make the following remarks:
\begin{itemize}
    \item[-] We consider the first day to be the observed stream prefix $S_0$ and use (part of) the queries $u \in \mathcal{U}_0' \subseteq \mathcal{U}_0$ therein (along with their number of occurrences during the first day) to learn the optimal hashing scheme via Algorithm \ref{alg:bcd} and for $\lambda=1$.
    \item[-] The first day consists of over 200,000 unique queries and just storing their IDs would require 200,000 buckets. Thus, we randomly sample a subset of the observed queries, with probabilities proportional to their observed frequencies. We use the sampled subset of queries as input to Algorithm \ref{alg:bcd}.
    \item[-] For fixed number of total buckets $b_{\text{total}}$, we need to determine the ratio $c$ between the number of buckets $b$ that the learned hashing scheme will consist of and the number of queries $n$ whose IDs we will store. Therefore, for user-specified $b_{\text{total}}$ and $c$, we pick $b$ and $n$ according to $$n = \nicefrac{b_{\text{total}}}{1+c},\quad b = b_{\text{total}}-n.$$
    In our experiments, we examine $c\in\{0.03,0.3\}.$
    \item[-] For the classifier $g: \mathcal{X} \rightarrow [b]$, mapping unseen queries $u \in \mathcal{U} \setminus \mathcal{U}_0'$ to buckets (as per Section \ref{sec:similarity-based-estimation}), we found that \rf\ achieves the best trade-off between training time and classification accuracy and use this model in the results we report.
    \item[-] To create input features for the classifier $g$, we follow a simple bag-of-words approach and only keep the $500$ most common words in the training queries. We also include as features the number of ASCII characters in the query text, the number of punctuation marks, the number of dots, and the number of whitespaces. As a result, the proposed approach is simple and interpretable, yet strong (as we show next).
\end{itemize}

\subsection{Results}

\begin{figure*}[!th]
\centering
\subfloat[Average per element absolute error after the $30^{th}$ day.]{\includegraphics[width=0.48\columnwidth]{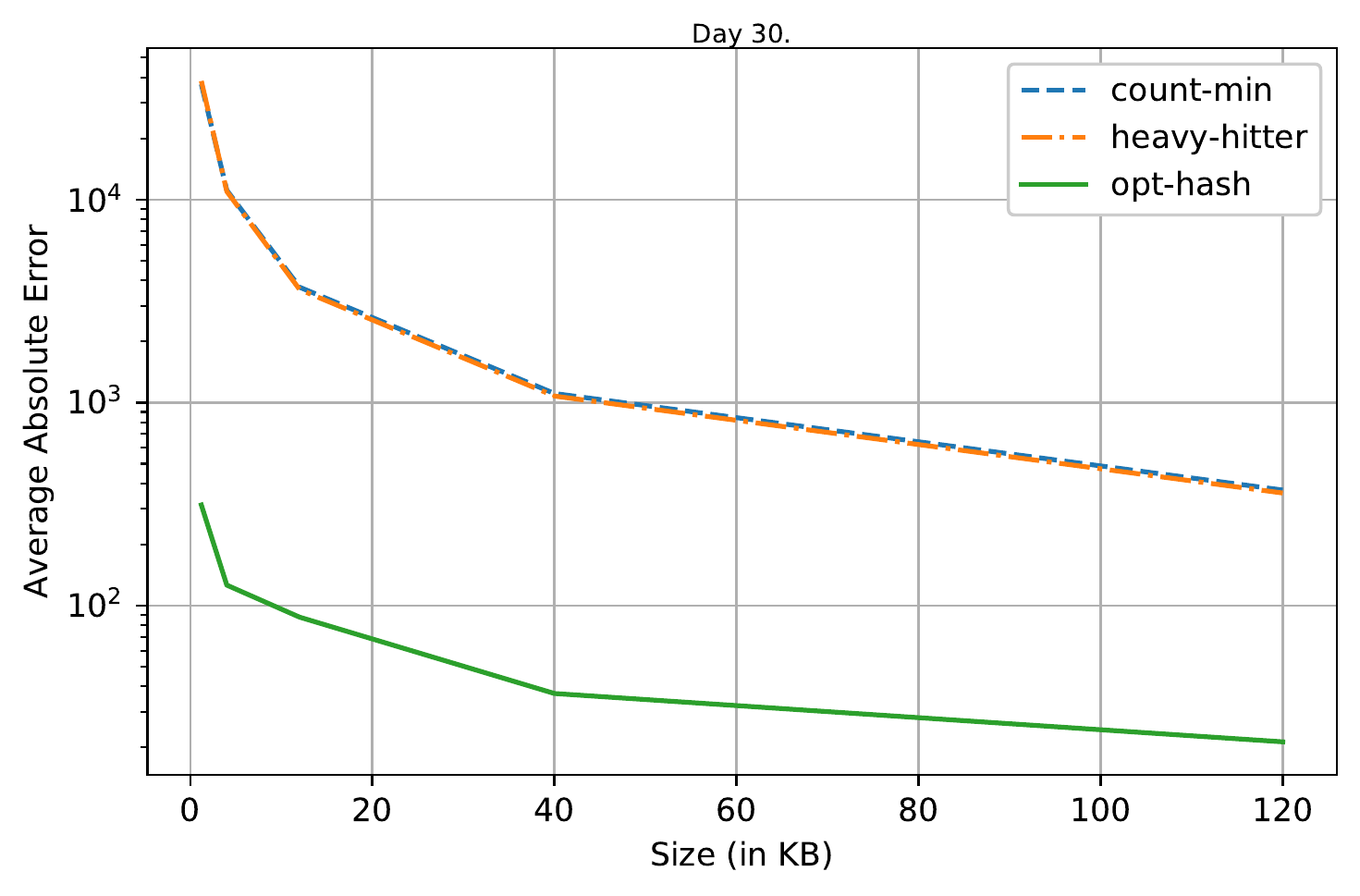}%
\label{fig:size_case_1}}
\hfil
\subfloat[Expected magnitude of absolute error after the $30^{th}$ day.]{\includegraphics[width=0.48\columnwidth]{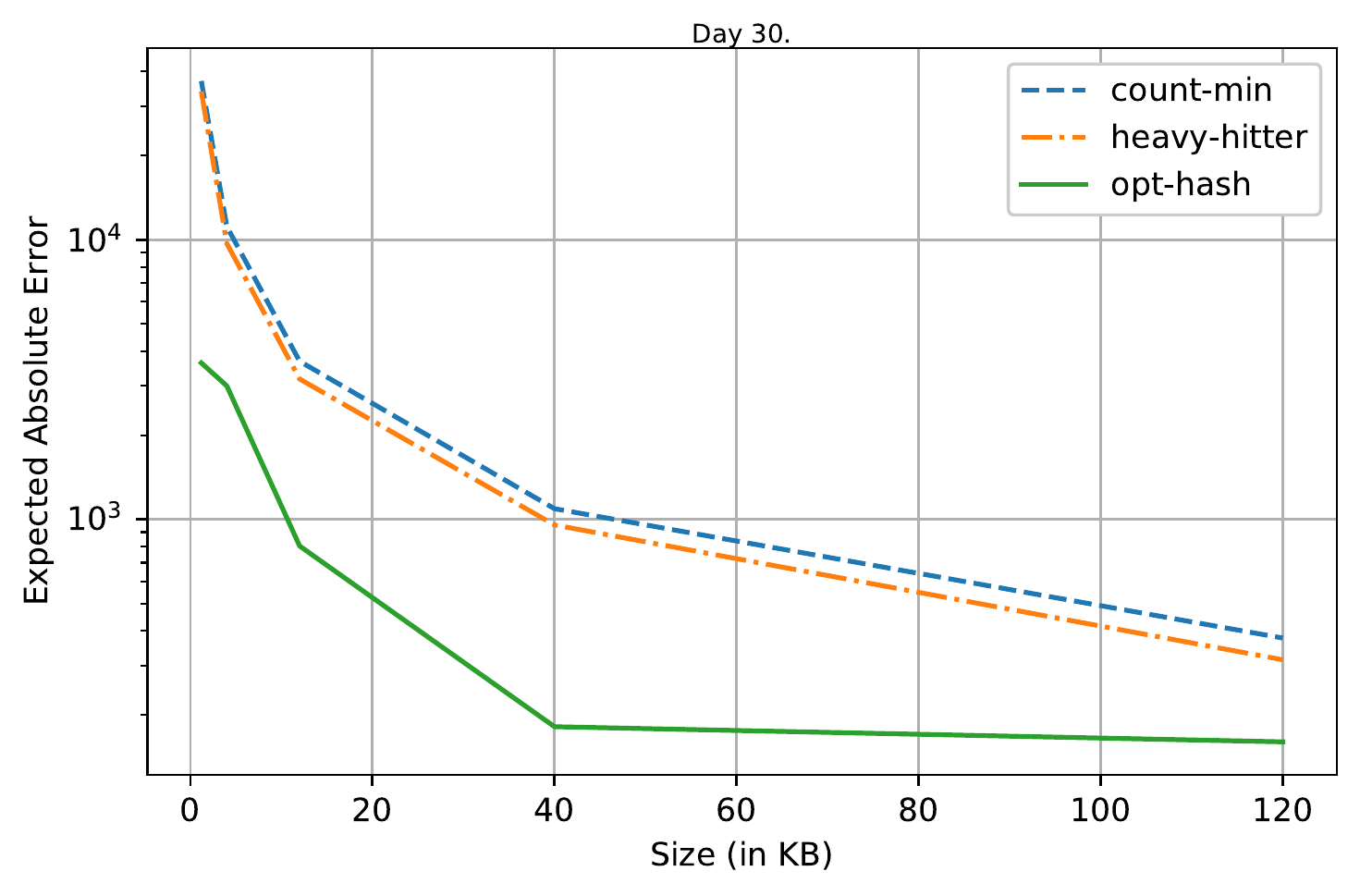}%
\label{fig:size_case_2}}
\hfil
\subfloat[Average per element absolute error after the $70^{th}$ day.]{\includegraphics[width=0.48\columnwidth]{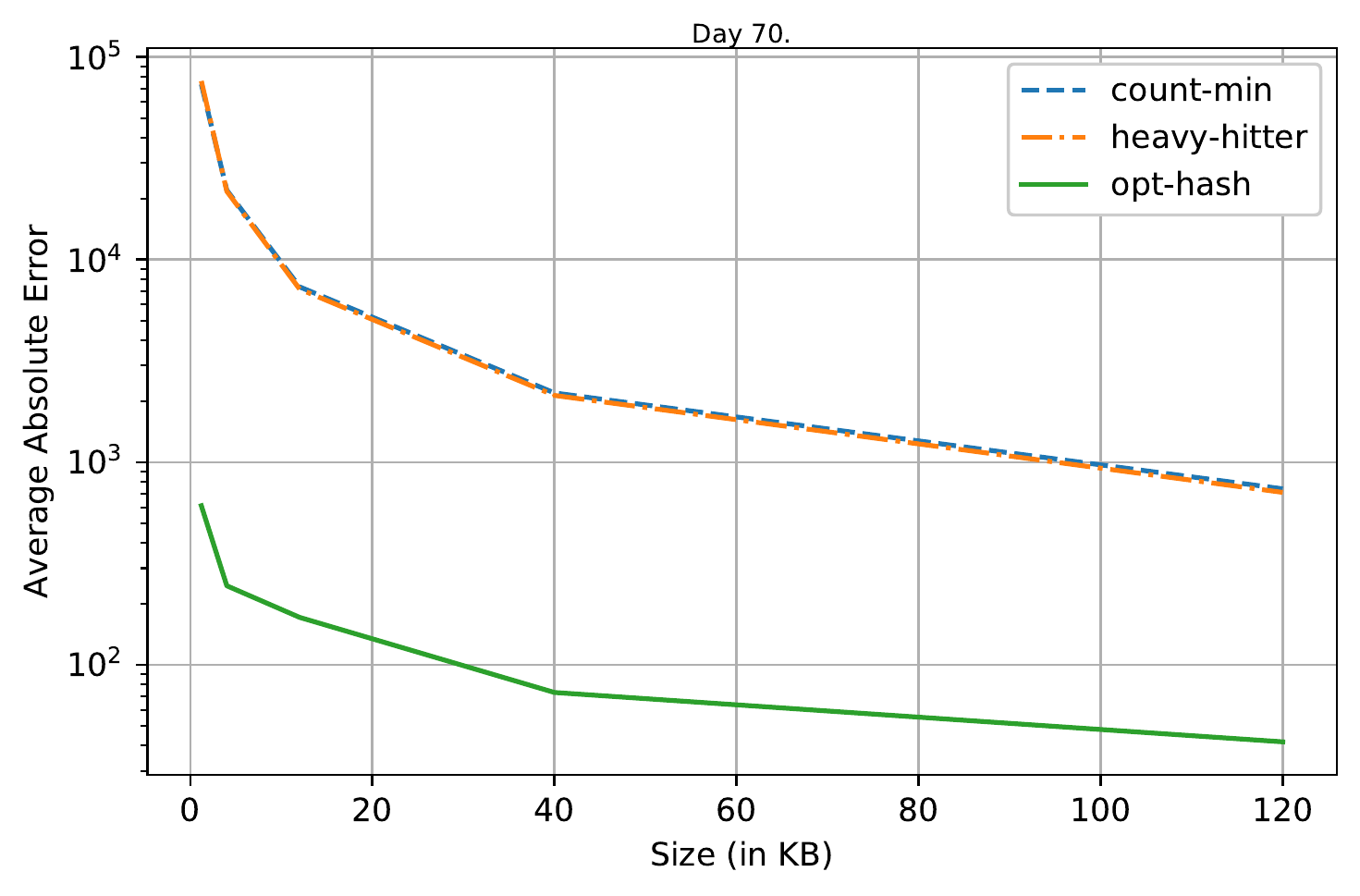}%
\label{fig:size_case_3}}
\hfil
\subfloat[Expected magnitude of absolute error after the $70^{th}$ day.]{\includegraphics[width=0.48\columnwidth]{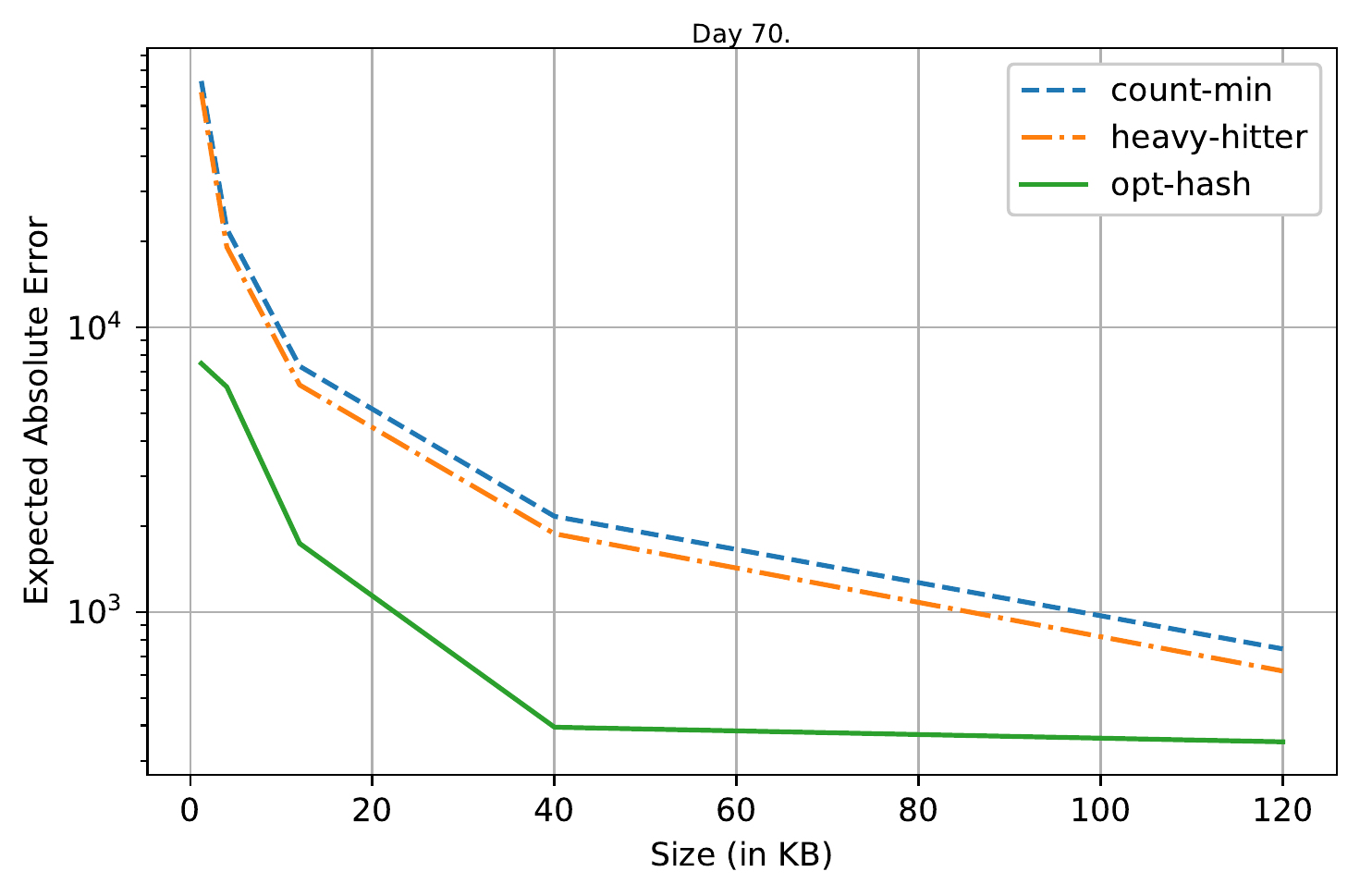}%
\label{fig:size_case_4}}
\caption{Estimation error as function of the estimator's size (in KB).}
\label{fig:size}
\end{figure*}

\begin{figure*}[!ht]
\centering
\subfloat[Average per element absolute error using 4 KB of memory. ]{\includegraphics[width=0.48\columnwidth]{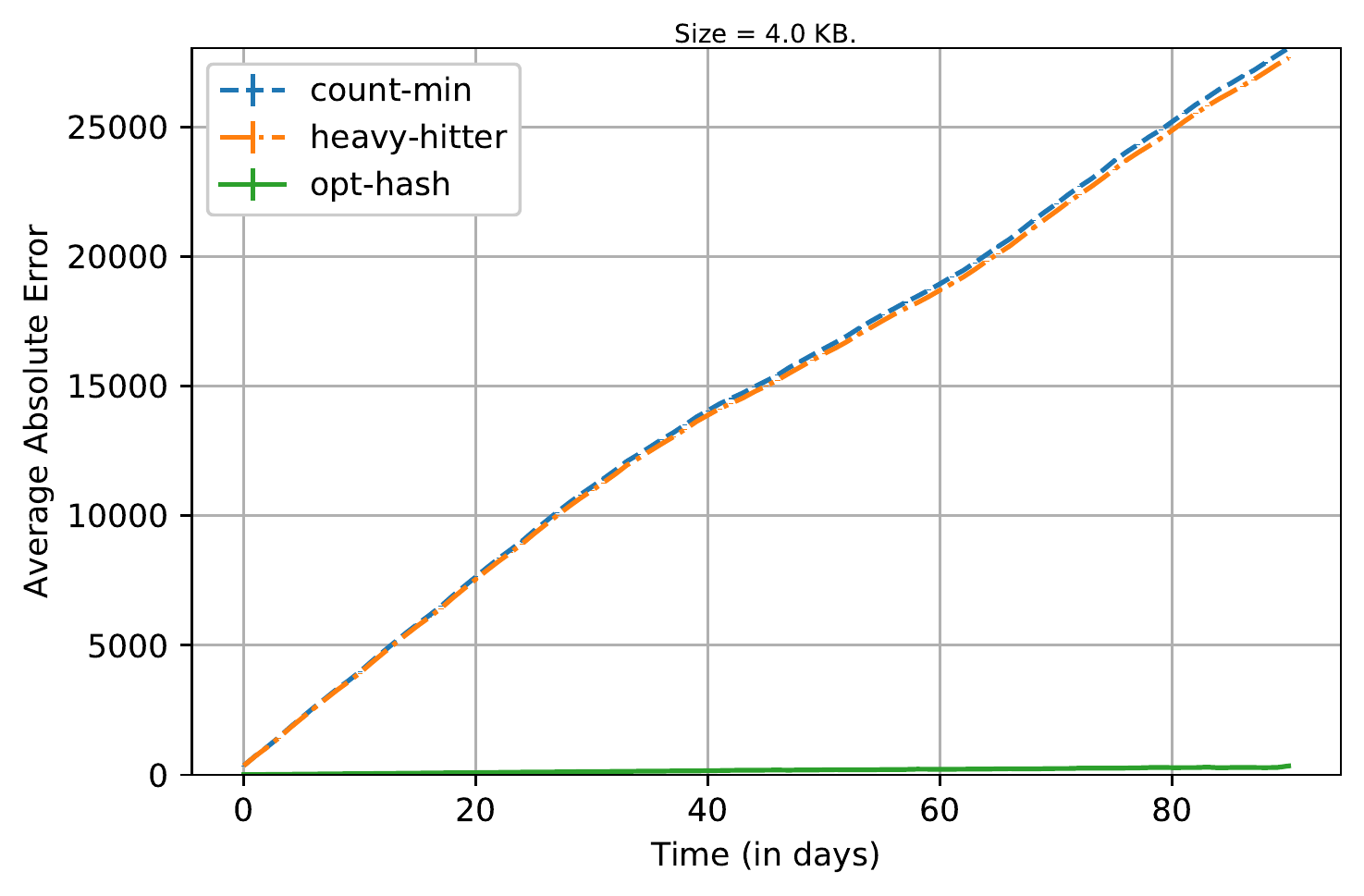}%
\label{fig:time_case_1}}
\hfil
\subfloat[Expected magnitude of absolute error using 4 KB of memory. ]{\includegraphics[width=0.48\columnwidth]{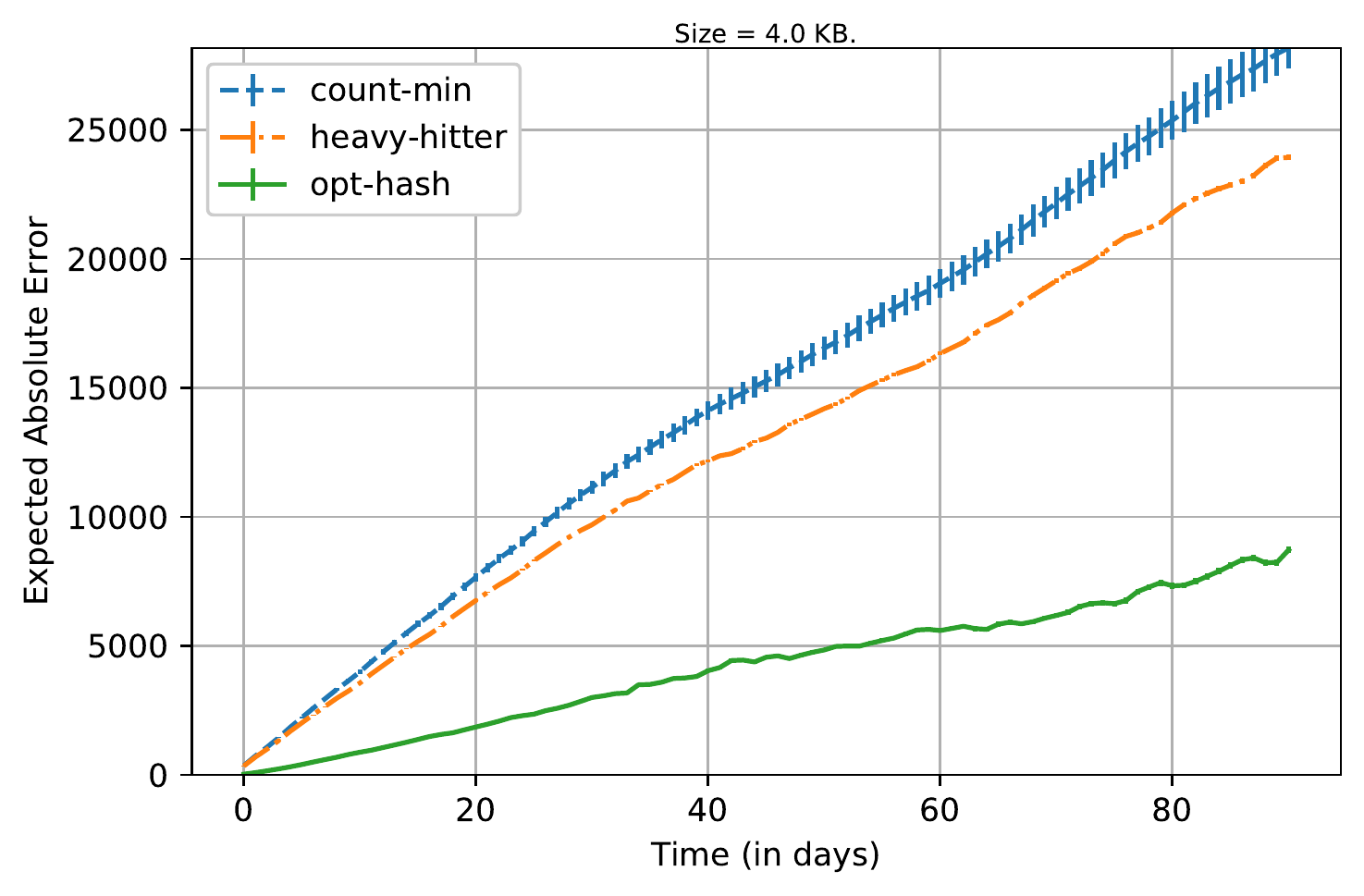}%
\label{fig:time_case_2}}
\hfil
\subfloat[Average per element absolute error using 120 KB of memory. ]{\includegraphics[width=0.48\columnwidth]{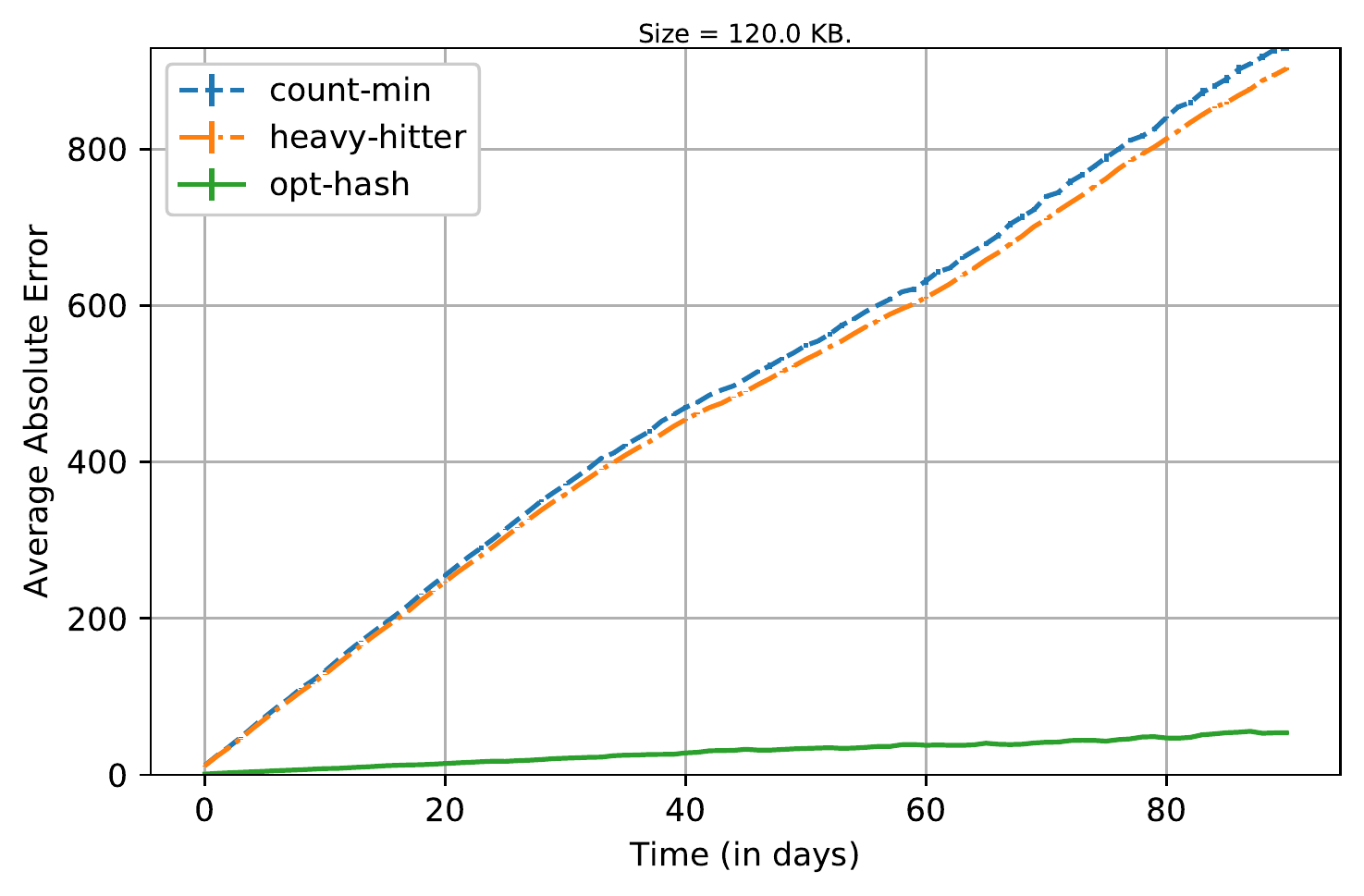}%
\label{fig:time_case_3}}
\hfil
\subfloat[Expected magnitude of absolute error using 120 KB of memory. ]{\includegraphics[width=0.48\columnwidth]{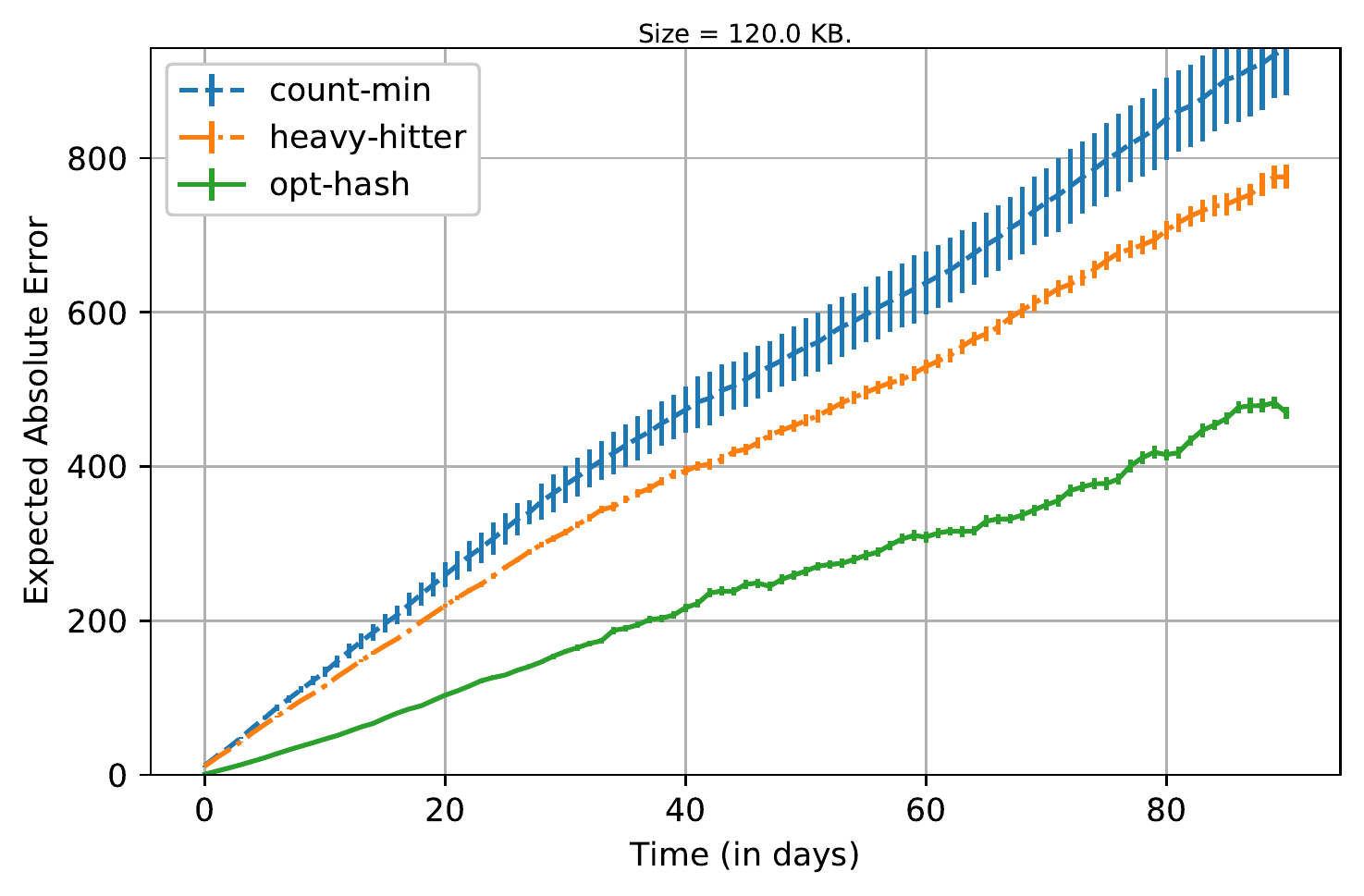}%
\label{fig:time_case_4}}
\caption{Estimation error as function of time (in days).}
\label{fig:time}
\end{figure*}

We implement our experiments in Python 3 and use the Scikit-learn machine learning package \cite{scikit-learn}. We independently repeat each experiment $5$ times and report the averaged error, as well as its standard deviation.
We remark that each bucket consumes $4$ bytes of memory and hence the total number of buckets used in each experiment can be calculated as $b=\frac{m\cdot10^3}{4},$ where $m$ is the size of the estimator in KB. Moreover, we denote by $\mathcal{U}_t$ the set of queries that appear in day $t$, and by $\*f_{\mathcal{U}_t}^t$ and $\*{\tilde{f}}_{\mathcal{U}_t}^t$ their aggregated true frequencies and estimated frequencies, respectively, between days $0$ and $t$.

In Figure \ref{fig:size}, we show the estimation error as function of the estimator's size in KB, after the $30^{th}$ and the $70^{th}$ day. On the the left (Figures \ref{fig:size_case_1} and \ref{fig:size_case_3}), we plot the average (per element) estimation error $$\frac{1}{|\mathcal{U}_t|} \sum_{u \in \mathcal{U}_t} |f_u^t-{\tilde{f}}_u^t|.$$ On the the right (Figures \ref{fig:size_case_2} and \ref{fig:size_case_4}), we plot the expected magnitude of the absolute estimation error $$\frac{1}{\sum_{u \in \mathcal{U}_t} f_u} \sum_{u \in \mathcal{U}_t} f_u^t \cdot |f_u^t-{\tilde{f}}_u^t|.$$ Notice that the former metric is expressed in a per element scale, that is, we normalize the overall error by the total number of elements $|\mathcal{U}_t|$ and hence all elements are penalized uniformly, whereas the second metric, the expected magnitude of the absolute estimation error, penalizes elements proportionally to their actual frequencies, as per Section \ref{sec:mip}.

We observe that the trend in the estimation error is very similar after the $30^{th}$ and the $70^{th}$ day. What changes is the absolute value of the estimation error, which, as expected, deteriorates with time, uniformly for all methods. The proposed method \oh\ consistently outperforms its competitors, in terms of both metrics. Unsurprisingly, as the size of all estimators increases, their errors drop. This is the case with both the average and the expected estimation error. We make the following additional remarks:
\begin{itemize}
    \item[-] The superiority of \oh\ is most notable in terms of average (per element) error. This is partly due to the fact that \oh\ does a substantially better job at estimating the frequencies of rarely occurring queries. In particular, queries that appear very few times are placed in the same bucket and hence the estimation error on them is small. In contrast, \hh\ and \cm\ often place such queries in the same bucket with queries of medium or even high frequencies, which produces big estimation error.
    \item[-] The expected magnitude of the estimation error of \hh\ and \cm\ does seem to slowly converge towards that of \oh\ when the estimators' size becomes sufficiently large. This indicates that \oh\ is particularly suited for low-space regimes and can achieve much more effective compression of the frequency vector.
    \item[-] As far as \hh\ and \cm\ are concerned, the former does produce better estimates, which is in agreement with the results in \cite{hsu2018learningbased}. The improvement is much more notable in terms of the expected magnitude of the estimation error. This observation is to be expected as well, given that \hh\ makes zero error on the most frequently occurring elements, which are heavily weighed in this metric.
\end{itemize}

Figure \ref{fig:time} reports the estimation error as function of time (in days), for two different memory configurations (4 KB in Figures \ref{fig:time_case_1} and \ref{fig:time_case_2}, 120 KB in Figures \ref{fig:time_case_3} and \ref{fig:time_case_4}). The superiority of \oh\ is preserved over time, in terms of both metrics. Moreover, we observe \oh\ achieves the smallest standard deviation in its estimation error. This can be attributed to the fact that the mappings of elements to buckets are more stable than those of \hh\ and \cm, as they are obtained via optimization instead of randomization; the main source of randomness for \oh\ is the classifier. 

We next experiment with memory configurations that vary between 1.2 KB and 120 KB, and compare \oh\ with \cm\ and \hh. The proposed approach provides an average improvement (over the entire 90-day period) by one to two orders of magnitude, in terms of its average (per element) absolute estimation error, and by 45-90\%, in terms of its expected magnitude of estimation error. For example, with 120 KB of memory, \oh\ makes an average absolute estimation error of $\sim29$ in estimating the frequency of each query, whereas the error of \hh\ is $\sim 479$ (Figure \ref{fig:size_case_1}). With 4 KB of memory, the errors of \oh\ and \hh\ are $\sim 167$ and $\sim14,661$, respectively (Figure \ref{fig:size_case_3}). {\color{edit_revision_1} Table \ref{tab:error} shows the average (per element) error after the entire 90-day period as a percentage of each query's frequency for the $1^{st}$, the $10^{th}$, the $100^{th}$, the $1,000^{th}$, and the $10,000^{th}$ most common queries.

\begin{table}[!ht]
\resizebox{\columnwidth}{!}{
\begin{tabular}{|l|l|l|}
\hline
\textbf{Query rank (by frequency)} & \textbf{Query frequency} & \textbf{Average error   percentage (\%)} \\ \hline
1 & 251,463 & 0.01 \\ \hline
10 & 37,436 & 0.08 \\ \hline
100 & 5,237 & 0.55 \\ \hline
1,000 & 926 & 3.13 \\ \hline
10,000 & 146 & 19.86 \\ \hline
\end{tabular}
}
\caption{Average (per element) error as percentage of query's frequency.}
\label{tab:error}
\end{table}
}

An additional feature of \oh\ is that, by using interpretable features in its machine learning component, it provides insights into the underlying frequency estimation problem. In particular, the features that were consistently marked as most important are the four counts (i.e., number of ASCII characters in the query text, the number of punctuation marks, the number of dots, and the number of whitespaces), as well as the words ``com,'' ``www,'' ``google,'' and ``yahoo.'' Intuitively, this observation makes sense. For instance, a large number of ASCII characters and whitespaces would be indicative of a big query with multiple words, making it more likely to be rare. On the other hand, a query containing the word ``google'' would be more likely to be common, given that ``google'' is consistently part of the most frequently occurring queries. 

\section{Conclusion} \label{sec:conclusion}
In this paper, we developed a novel approach for the problem of frequency estimation in data streams that relies on the use of optimization and machine learning on an observed stream prefix.
First, we formulated and efficiently solved the problem of optimally (or near-optimally) hashing the elements seen in the prefix to buckets, hence providing a smart alternative to oblivious random hashing schemes.
{\color{edit_revision_1}
To this end, we reformulated the problem as a mixed-integer linear optimization problem, we developed an efficient block coordinate descent algorithm, and, in a special case, we used dynamic programming to solve the problem in linear time.}
Next, we trained a classifier mapping unseen elements to buckets.
As we discussed, during stream processing, we only keep track of the frequencies of those elements that appeared in the prefix; the estimate the frequency of any element (either seen or unseen) is the average of the frequencies of all elements that map to the same bucket. 
We also described an adaptive approach that enables us to update the compressed frequency vector and keep track of the frequencies of all elements.
{\color{edit_revision_1}
We used synthetic data to investigate the performance, the scalability, and the impact of various design choices for the proposed approach; our study suggested that the proposed algorithms can compute optimal hashing schemes for problems with thousands of elements, using the mixed-integer linear optimization reformulation or the dynamic programming approach, and high quality hashing schemes for problems with tens of thousands of elements using the block coordinate descent algorithm.
}
Finally, we applied the proposed approach to the problem of search query frequency estimation and evaluated it using real-world data and empirically showed that the proposed learning-based streaming frequency estimation algorithm achieves superior performance compared to existing streaming frequency estimation algorithms.

\appendices

\section{Proof of Theorem \ref{theo:equiv}} \label{sec:appendix-proof}
\begin{proof}
We introduce variables $E \in \mathbb{R}_{\geq0}^{n \times b}$ such that $e_{ij}$ corresponds to the absolute estimation error associated with mapping element $i$ to bucket $j$. Since we are minimizing a nonnegatively weighed sum of such nonnegative terms, it suffices to require that
\begin{equation} \label{eqn:milp-pf-eq1}
    e_{ij} \geq f_i^0 - \frac{\sum_{k \in [n]} z_{kj} f^0_k}{\sum_{k \in [n]} z_{kj}},
    \qquad
    e_{ij} \geq - f_i^0 + \frac{\sum_{k \in [n]} z_{kj} f^0_k}{\sum_{k \in [n]} z_{kj}},
\end{equation}
for all $i \in [n], j \in [b]$.
To get rid of the fractional term in \eqref{eqn:milp-pf-eq1}, we multiply both equations with $\sum_{k \in [n]} z_{kj}$; this results in bilinear terms of the form $e_{ij} \sum_{k \in [n]} z_{kj}$. 
To linearize those, we introduce variables $\Theta \in \mathbb{R}_{\geq0}^{n \times n \times b}$ such that $\theta_{ikj} = e_{ij} z_{kj}$ can be interpreted as the error associated with mapping element $i$ to bucket $j$ when $k$ is also mapped therein. 
Since $\theta_{ikj}$ is the product of a binary variable and a continuous variable, we can linearize the constraint $\theta_{ikj} = e_{ij} z_{kj}$ by introducing a big-M constant such that, for all $i \in [n],j \in [b]$, $e_{ij} \leq M$.
We then require that
\begin{equation} \label{eqn:milp-pf-eq2}
    \theta_{ikj} \geq e_{ij} - M (1-z_{kj}), 
    \quad \theta_{ikj} \leq e_{ij},
    \quad \theta_{ikj} \leq M z_{kj},
\end{equation}
for all $i \in [n], k \in [n], j \in [b]$.
Thus, \eqref{eqn:milp-pf-eq1} can be rewritten as 
\begin{equation} \label{eqn:milp-pf-eq3}
\begin{split}
    &\sum_{k \in [n]} \theta_{ikj} \geq f_i^0 \sum_{k \in [n]} z_{kj} - \sum_{k \in [n]} f_k^0 z_{kj}, \\
    &\sum_{k \in [n]} \theta_{ikj} \geq - f_i^0 \sum_{k \in [n]} z_{kj} + \sum_{k \in [n]} f_k^0 z_{kj},
\end{split}
\end{equation}
which is linear in all variables.
To linearize the other bilinear term that appears in the objective function, we introduce another set of auxiliary variables $\Delta \in [0,1]^{n \times n \times b}$ such that $\delta_{ikj} = z_{ij} z_{kj}$ indicates whether elements $i$ and $k$ are mapped together to bucket $j$.
We then have the constraints
\begin{equation} \label{eqn:milp-pf-eq4}
    \delta_{ikj} \geq z_{ij} + z_{kj} - 1,
    \quad \delta_{ikj} \leq z_{ij},
    \quad \delta_{ikj} \leq z_{kj},
\end{equation}
for all $i \in [n], k \in [n], j \in [b]$.
Using the above new variables, the objective function can be written as
\begin{equation} \label{eqn:milp-pf-eq5}
    \sum_{i \in [n]} \sum_{j \in [b]} \left[ \lambda \theta_{iij} + (1-\lambda) \sum_{k \in [n]} \delta_{ikj} \| \*x_i - \*x_k\|^2 \right].
\end{equation}
Finally, we have to properly select the constant $M$ in \eqref{eqn:milp-pf-eq2} so that it is a valid upper bound for the variables $E$; such a bound can be obtained by setting $M \geq \max_{i \in [n]} f_i^0$, i.e., the estimation error associated with any element cannot be greater than the largest frequency observed in the prefix. 
\end{proof}

\section{Flowcharts for the Proposed Approach} \label{sec:appendix-flowchart}

\begin{figure*}[!ht]
\centering
\subfloat[Learning the optimal hashing scheme. ]{\includegraphics[width=0.98\columnwidth]{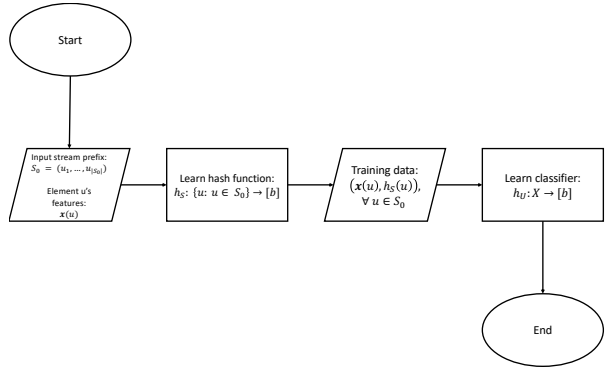}%
\label{fig:learn}}
\hfil
\subfloat[Answering count queries for element $u \in \mathcal{U}$. ]{\includegraphics[width=0.98\columnwidth]{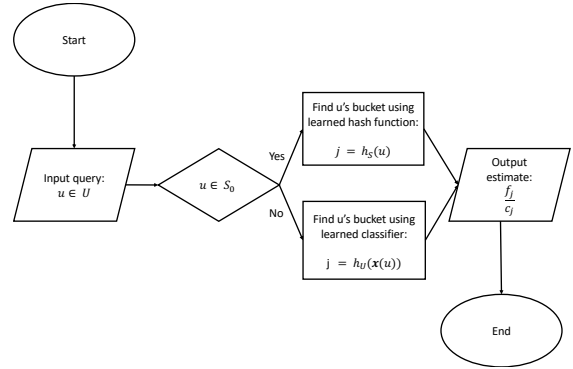}%
\label{fig:query}}
\hfil
\subfloat[Updating sketch at time $t$ upon arrival of element $u_t \in \mathcal{U}$. ]{\includegraphics[width=0.98\columnwidth]{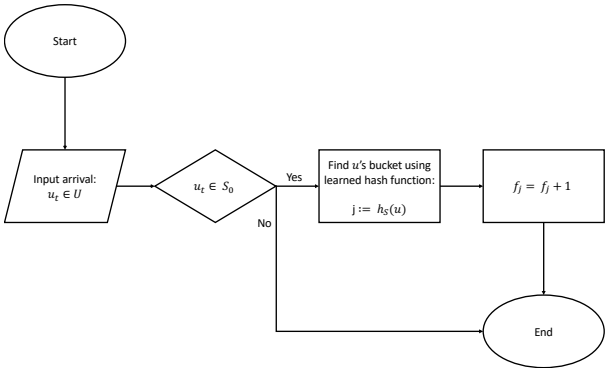}%
\label{fig:update}}
\hfil
\subfloat[Updating sketch with Bloom filter extension at time $t$ upon arrival of element $u_t \in \mathcal{U}$. ]{\includegraphics[width=0.48\textwidth]{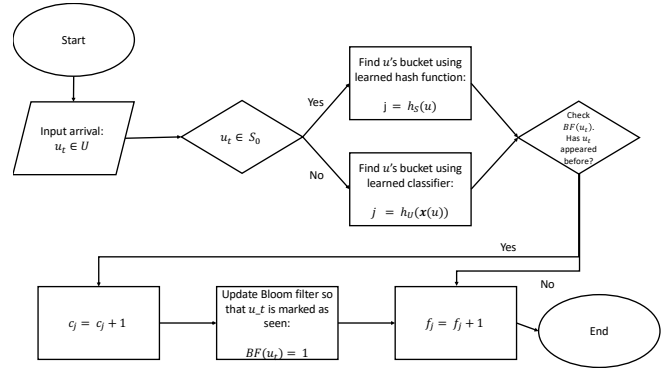}%
\label{fig:update-bloom}}
\hfil
\caption{Flowcharts for the proposed approach.}
\label{fig:flowcharts}
\end{figure*}

Figure \ref{fig:flowcharts} provides the flowcharts for the proposed approach. In particular, Figure \ref{fig:learn} corresponds to the learning phase, where the stream prefix is used to learn the optimal hashing scheme and the classifier; Figure \ref{fig:query} illustrates how the proposed approach answers count queries for any input element; Figures \ref{fig:update} and \ref{fig:update-bloom} show the update mechanism of the proposed approach without and with the use of Bloom filters, respectively.

\section{Table of Notations} \label{sec:appendix-notation}

Table \ref{tab:notations} includes the basic notations that are used repeatedly throughout the paper; we explain notations that are not used repeatedly in the main text. We note that we generally use the index $u$ to refer to elements, the indices $i$ and $k$ to refer to element IDs, the index $j$ to refer to buckets. For simplicity in notation, elements are referred to using either their symbol $u$ or their ID $i$, depending on the context; similarly, frequencies are indexed using either of the two approaches. 

\begin{table}[!ht]
\begin{tabularx}{\columnwidth}{|l|X|}
\hline
{\textbf{Symbol}} & {\textbf{Explanation}} \\ \hline
\multicolumn{2}{|l|}{\textit{General symbols:}} \\ \hline
{$\mathcal{U}$} & {Universe of elements} \\ \hline
{$\mathcal{U}_0$} & {Set of elements that appeared in the stream prefix} \\ \hline
{$n$} & {$|\mathcal{U}_0|$} \\ \hline
{$u \in \mathcal{U}$} & {Element} \\ \hline
{$k \in [|\mathcal{U}|]$} & {Element’s unique ID} \\ \hline
{$\mathcal{X}$} & {Feature space} \\ \hline
{$\boldsymbol{x} \in \mathcal{X}$} & {Element’s features} \\ \hline
{$\mathcal{S} = (u_1, \dots, u_{|\mathcal{S}|})$} & {Data stream} \\ \hline
{$\mathcal{S}_0$} & {Data stream prefix} \\ \hline
{$f_u$} & {Frequency of element $u$ in $\mathcal{S}$} \\ \hline
{$f_u^0$} & {Frequency of element $u$ in $\mathcal{S}_0$} \\ \hline
{$\tilde{f}_u$} & {Estimate of frequency of element $u$ in $\mathcal{S}$} \\ \hline
{$b$} & {Sketch’s total buckets} \\ \hline
\multicolumn{2}{|l|}{\textit{Symbols related to CMS and LCMS:}} \\ \hline
{$w$ and $d$} & {Sketch width and depth} \\ \hline
{$\phi_j$ (or $\phi_j^l$)} & {Aggregate frequency in bucket $j$ (or bucket $j$ in level $l$); this is used in CMS and LCMS} \\ \hline
{$h_{HH}(\cdot)$} & {Classifier that decides whether element $u$ is a heavy hitter} \\ \hline
\multicolumn{2}{|l|}{\textit{Symbols related to the proposed approach:}} \\ \hline
{$\mathcal{I}_j$} & {Set of elements in bucket $j$} \\ \hline
{$c_j$} & {Number of elements in bucket $j$} \\ \hline
{$\mu_j$} & {Mean of frequencies of elements in bucket $j$} \\ \hline
{$\boldsymbol{z}_i$} & {One-hot binary hash code for element with ID $i$} \\ \hline
{$h_i$} & {Integer hash code for element with ID $i$} \\ \hline
{$\lambda$} & {Hyperparameter that controls the trade-off between estimation error and similarity error} \\ \hline
{$h_S(\cdot)$} & {Function that maps elements that appeared in the prefix to   buckets based on the learned hash code} \\ \hline
$h_U(\cdot)$ & Classifier that maps elements to buckets \\ \hline
\end{tabularx}
\caption{Notations.}
\label{tab:notations}
\end{table}



\ifCLASSOPTIONcaptionsoff
  \newpage
\fi

\bibliographystyle{IEEEtran}
\bibliography{IEEEabrv,references}

\begin{IEEEbiography}[{\includegraphics[width=1in,height=1.25in,clip,keepaspectratio]{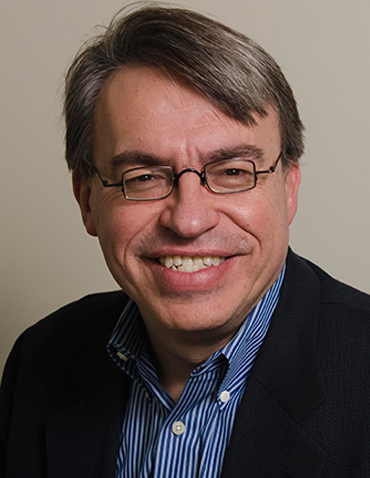}}]{Dimitris Bertsimas}
is the Associate Dean of Business Analytics, Boeing Professor of Operations Research and faculty director of the Master of Business Analytics at MIT. He received his SM and PhD in Applied Mathematics and Operations Research from MIT in 1987 and 1988 respectively. He has been MIT faculty since 1988. His research interests include optimization, machine learning, and applied probability, and their applications in health care, finance, operations management, and transportation. He has co-authored more than 200 scientific papers and five graduate level textbooks and has received numerous awards, with the most recent being the John von Neumann Theory Prize, INFORMS, and the President’s award, INFORMS, both in 2019.
\end{IEEEbiography}

\begin{IEEEbiography}[{\includegraphics[width=1in,height=1.25in,clip,keepaspectratio]{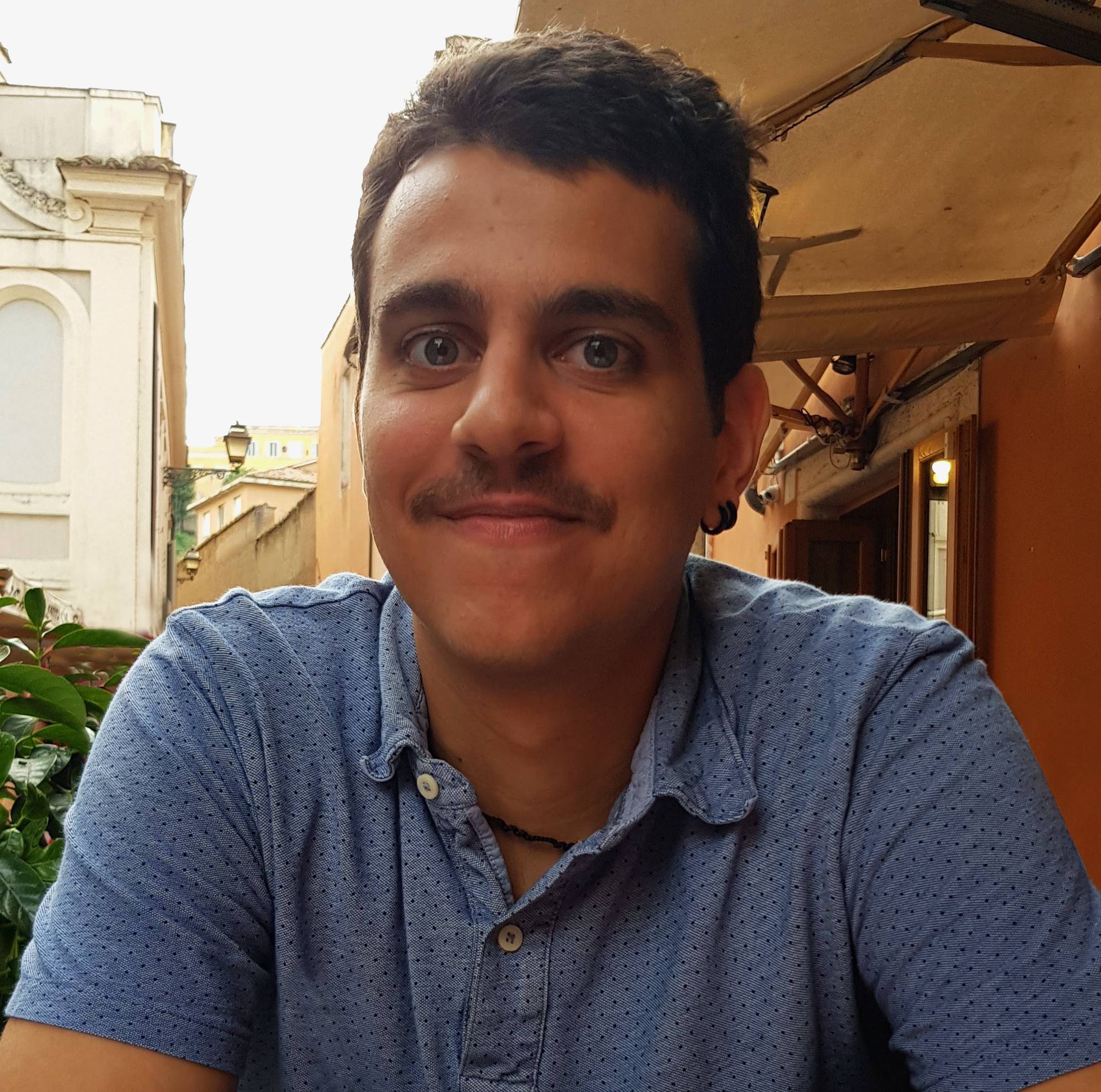}}]{Vassilis Digalakis Jr.}
is a PhD candidate at MIT’s Operations Research Center, advised by Prof. Dimitris Bertsimas. Prior to joining MIT, he earned his Diploma in Electrical and Computer Engineering from the Technical University of Crete, Greece, in 2018.
His research interests lie at the intersection of machine learning and optimization, with application to big-data settings. 
\end{IEEEbiography}

\end{document}